\documentclass[english,aps,tightenlines,showpacs,showkeys,notitlepage]{revtex4-2}
\usepackage[T1]{fontenc}
\usepackage[latin9]{inputenc}
\usepackage{color}
\usepackage{babel}
\usepackage{amsmath}
\usepackage{amssymb}
\usepackage{setspace}
\usepackage{esint}
\usepackage[unicode=true,pdfusetitle,
 bookmarks=true,bookmarksnumbered=false,bookmarksopen=false,
 breaklinks=false,pdfborder={0 0 1},backref=false,colorlinks=true]
 {hyperref}
\hypersetup{
 pdfborderstyle=}

\makeatletter
\usepackage{babel}

\makeatother

\begin{document}
\title{Generalized Brans-Dicke theory from Verlinde\textquoteright s entropic
gravity}
\author{Salih Kibaro\u{g}lu$^{1}$}
\email{salihkibaroglu@maltepe.edu.tr}

\author{Mustafa Senay$^{2}$}
\email{msenay@bartin.edu.tr}

\date{\today}
\begin{abstract}
In this work, we develop a $q$-deformed scalar-tensor theory of gravitation
by combining Verlinde\textquoteright s entropic gravity paradigm with
statistical deformation effects. The resulting model modifies the
Brans-Dicke framework through a deformation function, treated as a
constant rescaling factor for the effective gravitational coupling.
We derive the corresponding $q$-deformed field equations and analyze
their theoretical consistency including the recovery of standard gravitational
models in specific limits. While the present formulation preserves
key symmetries and provides a generalized description of gravitational
dynamics, it does not yield a significant deviation from Brans-Dicke
theory. The study concludes with prospects for future research, including
the exploration of cosmological solutions arising from the $q$-deformed
field equations.
\end{abstract}
\affiliation{$^{1)}$Maltepe University, Faculty of Engineering and Natural Sciences,
34857, Istanbul, Türkiye}
\affiliation{$^{2)}$Department of Medical Services and Techniques, Vocational
School of Health Services, Bart\i n University, 74100, Bart\i n, Türkiye}
\maketitle

\section{Introduction}

Einstein's theory of general relativity (GR) remains one of the most
successful and well-established theories in modern physics, offering
a robust framework for describing gravitational interactions and the
large-scale structure of the universe. However, despite these successes,
GR faces challenges in explaining the universe's accelerated expansion,
as well as the nature of dark energy and dark matter. To address these
issues, modified gravity theories have been developed that extend
GR by introducing additional fields or by modifying the space-time
geometry (see, e.g., \citep{Capozziello:2011Extended,Clifton:2011Modified,Nojiri:2017Modified}).
Among these, scalar-tensor theory stands out as a prominent example,
incorporating a scalar field that interacts with the metric, thereby
introducing additional degrees of freedom. This framework provides
a more versatile description of the universe\textquoteright s evolution,
particularly in explaining accelerated expansion without the necessity
of a cosmological constant.

Moreover, the principle of scale invariance \citep{Dirac:1936Wave,Mack:1969Finite}
plays a pivotal role in scalar-tensor theories, enabling the description
of phenomena without reference to an absolute scale. This principle
offers significant insights into both early-universe inflation and
late-time cosmic acceleration. By adapting to different cosmological
epochs, scalar-tensor theories effectively address several fundamental
challenges in cosmology \citep{Shaposhnikov:2009Scale,Israelit:2011Weyl,Bars:2013LocalConf,Aguila:2014PresentAccelerated,Karananas:2016Scale,Ferreira:2018Inflation,Casas:2019Scale,Ghilencea:2021GaugingScale,Tang:2020WeylSymmetry}.

Additionally, the discovery of Hawking radiation revealed that black
holes are thermal objects with entropy proportional to their horizon
area, known as the Bekenstein-Hawking entropy \citep{Bekenstein1973,Hawking:1975vcx},
thereby highlighting a fundamental link between gravity and thermodynamics
\citep{Jacobson:1995Thermodynamics,Padmanabhan:2005Gravity,Padmanabhan:2010Thermodynamical}.
Extending thermodynamic principles to the Hubble horizon has shown
that the FLRW equations can emerge from the first law of thermodynamics
\citep{Cai:2005FirstLaw,Akbar:2007Thermodynamic,Cai:2007Unified,Jamil:2010Thermodynamics,Sheykhi:2010Thermodynamics,Nojiri:2022Early,Nojiri:2022Modified,Nojiri:2024Different}.
In a similar vein, Verlinde's entropic gravity approach reinterprets
gravity as an emergent force arising from the statistical behavior
of microscopic degrees of freedom encoded in spacetime \citep{Verlinde:2010hp,Verlinde:2016toy}.
By associating entropy with holographic screens and viewing gravitational
interactions as changes in entropy due to matter displacements, Verlinde's
framework offers a thermodynamic perspective on gravity (for recent
developments, see \citep{Sheykhi:2010wm,Sheykhi:2010yq,Cai:2010hk,Cai:2010kp,Cai:2010zw,Wei:2010am,Sheykhi:2012vf,Komatsu:2012zh,Komatsu:2013qia,Moradpour:2018Implications,Senay:2018xaj,Kibaroglu:2018mnx,Moradpour:2019GupEup,Kibaroglu:2019dwd,Kibaroglu:2019odt,SENAY2021136536,Senay_2021,Senay_2024,Senay:2024Implications,Moradpour:2024Fractional,Moradpour:2024Tsallisian,Kibaroglu:2025Anisotropic}).
As demonstrated in these studies, such models provide a promising
framework for addressing several key cosmological challenges, as previously
mentioned.

In recent years, significant interest has been shown in understanding
the quantum properties of black holes. According to Strominger's proposal
\citep{Strominger:1993si}, quantum black holes can be considered
as obeying deformed Bose or Fermi statistics instead of standard Bose
or Fermi statistics. In this framework, quantum black holes can be
modeled as deformed particles, leading to profound for their statistical
and thermodynamic behavior. Deformed states of Bose and Fermi statistics
can be constructed by various approaches, one of which involves modifying
the quantum algebras by adding one or two deformation parameters \citep{Arik:1973vg,Parthasarathy_1991,Viswanathan_1992,Chaichian_1993}.
Extensive research has been conducted on this topic, and a large number
of studies have focused on the statistical and thermodynamic properties
of fermions and bosons in the context of deformed quantum algebras
\citep{Ubriaco1997,LAVAGNO2002310,Algin:2012df,Algin:2016cuo,Algin:2016df,Algin2015,Algin2017}.
These deformations not only provide deeper insights into the quantum
mechanical nature of black holes but also have broad applications
in various areas of physics. From quantum thermodynamics to cosmology,
deformed systems have been applied to a wide range of research areas,
highlighting their importance in understanding both microscopic and
macroscopic phenomena. The effects of deformation parameters on thermodynamic
curvature and Debye model heat capacity have been in the context of
a two-parameter deformed system \citep{Mohammadzadeh_2017}. A single-particle
$q$-deformed quantum harmonic oscillator has been studied to explore
how deformation affects its statistical complexity \citep{NUTKU2019122041}.
Considering the entropic gravitational framework and Strominger's
proposal about quantum black holes, the effects of fermionic and bosonic
$q$-deformation on the Einstein field equation and Friedmann equations
have been investigated \citep{Senay:2018xaj,Kibaroglu:2018mnx,Kibaroglu:2019odt,SENAY2021136536}.
By applying $q$-deformed fermionic statistics to the degrees of freedom
in the holographic screen and using Verlinde's hypothesis, it has
been shown that the MOND interpolation function can be interpreted
as the heat capacity per degree of freedom of the holographic screen
\citep{Senay_2021}. Using Schwinger notation and $q$-deformed harmonic
oscillators, a three-level quantum system (qutrits) has been constructed
along with $q$-deformed quantum logic gates for one-, two- and three-qutrit
systems \citep{Altintas:2020eqx}. A quantum Otto cycle has been studied
using a $q$-deformation quantum oscillator and classical thermal
baths, and how the quantum statistical deformation parameter $q$
affects the work output and efficiency of the cycle has been analyzed
\citep{Ozaydin:2023ege}. The effects of the deformation parameter
on the thermal and electrical conductivities of some elements as a
function of the deformed Debye specific heat have been investigated
by considering the Einstein and Debye models \citep{Marinho2024}.
Using the generalized Duffin-Kemmer-Petiau equation, which includes
a new non-minimal coupling based on the $q$-deformed formalism, the
thermodynamic properties and relativistic behavior of a spin-one boson
particle in one-dimensional space have been investigated \citep{BOUMALI2023129134}.
Additionally, deformed systems have found applications in a wide range
of research fields across physics and beyond \citep{Senay_2024,Boutabba2022,Nutku_2018,MARINHO201474,Gavrilik2014,ISMAIL2003259,JALALZADEH2023101320}.
These applications demonstrate the versatility and impact of deformed
systems in both theoretical and applied science.

Although Verlinde's entropic gravity model has been widely studied
in the literature, its application to scale-invariant gravitational
theories has not yet received sufficient attention. The main purpose
of this paper is to unify Verlinde's entropic gravity approach with
$q$-deformed statistics and the principle of scale-invariance in
the metric tensor. The paper is organized as follows: after providing
a short discussion of the deformed algebra and the thermodynamical
background under $q$-deformation in Section \ref{sec:Deformation-of-the},
we introduce the scalar-tensor theory of gravity formulated within
the framework of $q$-statistics in Section \ref{sec:Scalar-tensor-theory}.
The paper concludes with final remarks and potential avenues for future
research in Section \ref{sec:Conclusion}.

Notations and conventions: In this work, we adopt the mostly negative
signature for the metric, such that the four-dimensional Minkowski
metric takes the form $\text{diag}(1,-1,-1,-1)$. Throughout the paper,
we work in natural units by setting the fundamental physical constants
to unity: the speed of light $c=1$, Newton's gravitational constant
$G=1$, and Boltzmann\textquoteright s constant $k_{B}=1$. The four-dimensional
space-time indices are represented by Roman letters, such as $a,b,c,...=0,1,2,3$.

\section{Deformed algebra and thermodynamical background\label{sec:Deformation-of-the}}

This section explores the quantum algebraic structure and thermodynamic
behavior of the $q$-deformed fermion gas model, which extends the
conventional Pauli exclusion principle to provide deeper insights
into the physics of quasiparticle systems. The algebraic framework
for this model, pioneered in Refs. \citep{Parthasarathy_1991,Viswanathan_1992,Chaichian_1993},
represents a significant advancement in the understanding of quantum
deformation. Additionally, the high- and low-temperature thermodynamic
properties of the $q$-deformed fermion system, thoroughly analyzed
in Refs. \citep{Algin:2012df,Algin:2016cuo,Algin:2016df}, reveal
its unique statistical behavior. These properties are governed by
the following deformed anticommutation relations:
\begin{equation}
aa^{*}+qa^{*}a=1,\label{eq:q-algebra1}
\end{equation}
\begin{equation}
[\hat{N},a^{*}]=a^{*},\,\,\,\,[\hat{N},a]=-a.\label{eq:q-algebra2}
\end{equation}
Here, $a$ and $a^{*}$ denote the deformed fermion annihilation and
creation operators, respectively, while $\hat{N}$ represents the
fermion number operator. The parameter $q$, which characterizes the
degree of deformation, is a positive real number constrained to the
interval $0<q<1$. These relations reflect the modified statistical
behavior introduced by the $q$-deformation, which diverges from the
standard fermionic anticommutation relations.

The spectrum of the fermion number operator, $\hat{N}$, which determines
the permissible particle number states in this deformed framework,
is given by:

\begin{equation}
\left[n\right]=\frac{1-(-1)^{n}q^{n}}{1+q}.\label{eq:spectrum}
\end{equation}
The thermodynamic properties of the $q$-deformed fermion gas model
can be systematically studied using the fermionic Jackson derivative
operator, which provides a generalized differentiation framework aligned
with the algebraic structure of the model. This operator replaces
the conventional derivative operator and is formally defined as
\begin{equation}
D_{x}^{q}f(x)=\frac{1}{x}\left[\frac{f(x)-f(-qx)}{1+q}\right],\label{eq:Jackson}
\end{equation}
where $f(x)$ represents any differentiable function. Within the framework
of the $q$-deformed fermion gas model, the average occupation number,
which describes the statistical distribution of particles over quantum
states, can be rigorously expressed as
\begin{equation}
n_{i}=\frac{1}{|\ln q|}\left|\ln\left(\frac{|1-ze^{-\beta\epsilon_{i}}|}{1+qze^{-\beta\epsilon_{i}}}\right)\right|,\label{eq:q-occupation}
\end{equation}
where $\epsilon_{i}$ represents the kinetic energy associated with
the $i$-th one-particle energy state, and $\beta=1/T$ is the inverse
thermal energy, with $T$ being the absolute temperature of the system.
The parameter $z=\exp\left(\mu/T\right)$ is the fugacity, defined
in its standard form, where $\mu$ is the chemical potential of the
system.

Utilizing Eq.(\ref{eq:q-occupation}), the total number of particles
$N$ and the internal energy $U$ of the $q$-deformed fermion gas
system are mathematically defined as follows

\begin{equation}
N=\sum_{i}\frac{1}{|\ln q|}\left|\ln\left(\frac{|1-ze^{-\beta\epsilon_{i}}|}{1+qze^{-\beta\epsilon_{i}}}\right)\right|,\label{eq:totalnumber1}
\end{equation}
and 
\begin{equation}
U=\sum_{i}\frac{\epsilon_{i}}{|\ln q|}\left|\ln\left(\frac{|1-ze^{-\beta\epsilon_{i}}|}{1+qze^{-\beta\epsilon_{i}}}\right)\right|,\label{eq:internalenergy1}
\end{equation}

In the thermodynamic limit, where the volume and the number of particles
in the system become large, the discrete summations over quantum states
in Eqs.(\ref{eq:totalnumber1}) and (\ref{eq:internalenergy1}) can
be replaced by continuous integrals over the phase space. Hence, the
following fundamental relations can be readily obtained:

\begin{equation}
\frac{N}{V}=\frac{1}{\lambda^{3}}f_{3/2}(z,q),\label{eq:totalnumber2}
\end{equation}
\begin{equation}
\frac{U}{V}=\frac{3}{2}\frac{T}{\lambda^{3}}f_{5/2}(z,q),\label{eq:internalenergy2}
\end{equation}
where $V$ denotes the system's volume, and $\lambda=h/(2\pi mT)^{1/2}$
represents the thermal de Broglie wavelength, with $h$ being Planck's
constant and $m$ the particle mass. These expressions are derived
by employing the density of states, given as $(V/2\pi^{2})(2m/\hbar^{2})^{3/2}\epsilon^{1/2}$,
where $\hbar=h/2\pi$ is the reduced Planck constant. The function
$f_{n}(z,q)$ is the generalized Fermi-Dirac function, which accounts
for the statistical behavior of the particles in the deformed system
which is defined as;

\begin{eqnarray}
f_{n}(z,q) & = & \frac{1}{\Gamma(n)}\int_{0}^{\infty}\frac{x^{n-1}}{|\ln q|}\left|\ln\left(\frac{|1-ze^{-x}|}{1+qze^{-x}}\right)\right|\text{d}x\nonumber \\
 & = & \frac{1}{|\ln q|}\left[\sum_{l=1}^{\infty}(-1)^{l-1}\frac{(zq)^{l}}{l^{n+1}}-\sum_{l=1}^{\infty}\frac{z^{l}}{l^{n+1}}\right],\label{eq:q-FD}
\end{eqnarray}
where $\Gamma\left(n\right)=\int_{0}^{\infty}x^{n-1}e^{-x}dx$ is
the Gamma function for $n>0$ and $x=\beta\epsilon$. From the thermodynamic
relation $F=\mu N-PV$, where $P=2U/3V$ is the pressure of the system,
the Helmholtz free energy can be expressed as
\begin{equation}
F=\frac{TV}{\lambda^{3}}\left[f_{3/2}(z,q)\ln z-f_{5/2}(z,q)\right].\label{eq:Helmholtz}
\end{equation}
The entropy function of the system is derived from the relation $S=(U-F)/T$,
leading to
\begin{equation}
S=\frac{V}{\lambda^{3}}\left[\frac{5}{2}f_{5/2}(z,q)-f_{3/2}(z,q)\ln z\right].\label{eq:q-entropy}
\end{equation}
Assuming the one-particle kinetic energy is given by $E=T$, Eq. (\ref{eq:q-entropy})
can be re-expressed in terms of the kinetic energy $E$ as
\begin{equation}
S=\frac{V(2\pi m)^{3/2}}{Th^{3}}E^{5/2}\left[\frac{5}{2}f_{5/2}(z,q)-f_{3/2}(z,q)\ln z\right],\label{eq:q-entropy1}
\end{equation}
This formulation provides a direct connection between the system's
entropy, the kinetic energy, and the generalized Fermi-Dirac function,
incorporating the effects of $q$-deformation on the thermodynamic
properties. 

We demonstrated key thermodynamic properties (such as particle number
density, internal energy, Helmholtz free energy, and entropy) highlighting
the impact of deformation on the system. The next section examines
how $q$-deformation affects gravitational dynamics under the scale
invariance of the metric tensor, linking quantum statistical mechanics
with gravitational fields.

\section{\label{sec:Scalar-tensor-theory}Scalar tensor theory of gravity
under $q$-statistics}

According to Verlinde's approach \citep{Verlinde:2010hp,Verlinde:2016toy},
gravity emerges as an entropic force when mass is distributed on a
holographic screen, consistent with the holographic principle. In
this section, we investigate how the $q$-deformed fermionic model
influences gravitational dynamics by analyzing its impact on the Einstein
field equations under a conformal transformation of the metric tensor.

We begin by adopting a natural generalization of the Newtonian gravitational
potential within the framework of general relativity \citep{Wald:1984GeneralRelativity}.
In a stationary space-time that admits a time-like Killing vector
field $\xi^{a}$, the relativistic gravitational potential $\phi$
is defined as

\begin{equation}
\phi=\frac{1}{2}\log\sqrt{-\xi^{a}\xi_{a}}\label{eq: grav_pot}
\end{equation}
where $\xi^{a}$ encodes the time-translation symmetry of the space-time.
For a static, asymptotically flat space-time (vacuum in the neighborhood
of spatial infinity), the Killing vector is normalized such that the
redshift factor, $\sqrt{-\xi^{a}\xi_{a}}\rightarrow1$, as $r\rightarrow\infty$.
Accordingly, the exponential factor $e^{\phi}$ represents the redshift
between local and asymptotic observers, with $\phi=0$ taken as the
reference point at spatial infinity.

In such a static space-time, the notion of an observer \textquotedbl at
rest\textquotedbl{} is well-defined and corresponds to worldlines
that are tangent to the timelike Killing vector field $\xi^{a}$.
Denoting the four-velocity of such an observer as $u^{a}$, the proper
acceleration $a^{b}=u^{b}\nabla_{a}u^{a}$ can be expressed in terms
of the Killing vector as follows:

\begin{equation}
u^{b}=e^{-\phi}\xi^{b},\,\,\,\,\,\,\,\,\,\,\,\,\,\,\,\,\,\,\,\,\,\,\,\,\,a^{b}=e^{-2\phi}\xi^{a}\nabla_{a}\xi^{b}.
\end{equation}
Using the Killing equation, $\nabla_{a}\xi_{b}+\nabla_{b}\xi_{a}=0$,
and the definition of the gravitational potential given in (\ref{eq: grav_pot}),
the acceleration simplifies to
\begin{equation}
a^{b}=-\nabla^{b}\phi.
\end{equation}
This expression shows that the gravitational acceleration, analogous
to the Newtonian case, is directed orthogonally to the equipotential
surfaces, which are identified in this context with holographic screens
$\mathcal{S}$. By contracting with a unit normal vector $N^{b}$,
orthogonal to both the screen $\mathcal{S}$ and the Killing vector
$\xi^{b}$, one finds that the local force required to keep a unit
test mass stationary is given by
\begin{equation}
F_{a}=-me^{\phi}\nabla_{a}\phi,\label{eq: force}
\end{equation}
where the additional term $e^{\phi}$ arises due to the redshift between
the local frame and the reference frame at infinity.

Within this geometric framework, we now incorporate the effects of
$q$-deformation by considering the modified entropy function introduced
in Eq.(\ref{eq:q-entropy1}). Following the formalism developed in
Refs. \citep{Senay:2018xaj,Kibaroglu:2018mnx,Kibaroglu:2019odt,Senay:2024Implications},
we derive the corresponding $q$-deformed temperature. At equilibrium,
the system maximizes its entropy, implying that the total entropy
$S$ remains stationary under an infinitesimal virtual displacement.
Mathematically, this condition is expressed as
\begin{equation}
\frac{\text{d}}{\text{d}x^{a}}S(E,x^{a})=0,\label{eq:sentropy}
\end{equation}
which leads to the differential relation

\begin{equation}
\frac{\partial S}{\partial E}\frac{\partial E}{\partial x^{a}}+\frac{\partial S}{\partial x^{a}}=0,\label{eq:sentropy1}
\end{equation}
Using the relations $\frac{\partial E}{\partial x^{a}}=-F_{a}$ and
$\frac{\partial S}{\partial x^{a}}=\nabla_{a}S$, we further impose
that the entropy change across an infinitesimal displacement of one
Compton wavelength normal to the screen satisfies:
\begin{equation}
\nabla_{a}S=-2\pi\frac{m}{\hbar}N_{a},\label{eq: change_in_S}
\end{equation}
where $N^{a}$ is the unit outward normal vector. 

Substituting Eqs.(\ref{eq: force}) and (\ref{eq: change_in_S}) into
(\ref{eq:sentropy1}), the local temperature $T$ on the holographic
screen, including the gravitational redshift correction, becomes:

\begin{equation}
T=\alpha\left(z,q\right)\frac{\hbar}{2\pi}e^{\phi}N^{a}\nabla_{a}\phi.\label{eq:q-temperature}
\end{equation}
Here, the redshift factor $e^{\phi}$ is included because the temperature
$T$ is measured relative to the reference frame at spatial infinity.
The function $\alpha\left(z,q\right)$ which encapsulates the effect
of the $q$-deformation and the fugacity parameter $z$, is given
by (see Appendix I):
\begin{equation}
\frac{\alpha\left(z,q\right)}{N}=\frac{5}{2}\left[\frac{5}{2}\frac{f_{5/2}(z,q)}{f_{3/2}(z,q)}-\ln z\right].\label{eq:alpha}
\end{equation}
Finally, the $q$-deformed temperature (\ref{eq:q-temperature}) can
be expressed compactly as:

\begin{equation}
T=\alpha\left(z,q\right)T_{U},
\end{equation}
where $T_{U}=\frac{\hbar}{2\pi}e^{\phi}N^{a}\nabla_{a}\phi$ is the
Unruh temperature \citep{Unruh:1976Notes,Verlinde:2010hp} which is
associated with the thermal radiation observed by an accelerated observer
in a gravitational field, and its dependence on the gravitational
potential $\phi$ reflects the influence of space-time geometry on
the temperature of the system. 

In scalar-tensor theory, two distinct frames are commonly used: the
Jordan frame and the Einstein frame, with their respective metrics
denoted by $g_{ab}$ and $\tilde{g}_{ab}$. These frames are related
through a conformal transformation, expressed as
\begin{equation}
\tilde{g}_{ab}=\psi g_{ab},\label{eq:metricT}
\end{equation}
where $\psi$ is a scalar field that non-minimally couples to the
scalar curvature in the Jordan frame. In the Einstein frame, the gravitational
potential is redefined as
\begin{equation}
\tilde{\phi}=\phi+\frac{1}{2}\log\psi.
\end{equation}
Using this conformal formalism, the temperature $T$ defined in (\ref{eq:q-temperature})
can be generalized to
\begin{equation}
T=\alpha\left(z,q\right)\frac{\hbar}{2\pi}e^{\phi}N^{a}\left(a_{a}+\frac{1}{2}\nabla_{a}\log\psi\right),\label{eq:q-temperature1}
\end{equation}
where $a_{a}=\nabla_{a}\phi$ is the acceleration measured in Jordan
frame. The additional term $\frac{1}{2}\nabla_{a}\log\psi$ accounts
for the effect of the scalar field $\psi$ in the Einstein frame,
modifying the system's temperature due to the presence of the non-minimal
coupling between the scalar field and the geometry of space-time. 

According to Bekenstein idea \citep{Bekenstein1973}, if a test particle
is located close to the event horizon of a black hole but far from
the Compton wavelength, this increases the mass of the black hole,
and this process is considered as a bit of information. Considering
a holographic screen on a closed surface with a constant redshift,
the differential of the number of bits can be expressed as: 
\begin{equation}
\text{d}N=\frac{\psi\text{d}A}{\hbar}.\label{eq: diff_number}
\end{equation}
Here $A$ represents the area of the closed surface on the screen.
Now, suppose that the total energy $E$ is related to the total mass
$M$ distributed among all the bits on the screen. Using the law of
equipartition of energy and the relation $E=M$, the total mass can
be expressed as 
\begin{equation}
M=\frac{1}{2}\oint_{\partial\Sigma}T\text{d}N.\label{eq:mass}
\end{equation}
Substituting Eqs. (\ref{eq:q-temperature1}) and (\ref{eq: diff_number})
into Eq. (\ref{eq:mass}), one can easily reach 
\begin{equation}
M=\frac{\alpha\left(z,q\right)}{4\pi}\oint_{\partial\Sigma}e^{\phi}N^{a}\left(a_{a}+\frac{1}{2}\nabla_{a}\log\psi\right)\psi\text{d}A.\label{eq:q-mass}
\end{equation}
This is called the $q$-deformed Gauss law under scalar-tensor gravity.
Using Stokes theorem in Eq. (\ref{eq:q-mass}), this equation can
be written as (for detailed calculation, see Appendix II)

\begin{equation}
M=\frac{\alpha\left(z,q\right)}{4\pi}\int_{\Sigma}\left[\psi R_{ab}-\left(\nabla_{a}\nabla_{b}+\frac{1}{2}g_{ab}\square\right)\psi\right]\xi^{b}\text{d}\Sigma^{a},\label{eq:q-mass1}
\end{equation}
where $\square=g^{ab}\nabla_{a}\nabla_{b}$ denotes the d'Alembert
operator and $R_{ab}$ is Ricci tensor. 

On the other hand, an alternative definition of the Komar mass is
provided by the following expression:

\begin{equation}
M=2\int_{\Sigma}\left(T_{ab}-\frac{1}{2}g_{ab}T_{c}^{c}\right)\xi^{b}\text{d}\Sigma^{a},\label{eq:komarmass}
\end{equation}
where $T_{ab}$ is the energy-momentum tensor. Comparing the Eq. (\ref{eq:q-mass1})
with Eq. (\ref{eq:komarmass}), the Ricci tensor can be obtained as
\begin{equation}
R_{ab}=\frac{8\pi}{\alpha\left(z,q\right)\psi}\left(T_{ab}-\frac{1}{2}g_{ab}T\right)+\frac{1}{\psi}\left(\nabla_{a}\nabla_{b}+\frac{1}{2}g_{ab}\square\right)\psi,\label{eq: Riccitensor1}
\end{equation}
where $T=g^{ab}T_{ab}$ is the trace of the energy-momentum tensor.
To find the corresponding Ricci scalar $R$, we take the trace of
the Ricci tensor $R_{ab}$ with the metric $g^{ab}$:
\begin{eqnarray}
R & = & -\frac{8\pi}{\alpha\left(z,q\right)\psi}T+\frac{3}{\psi}\square\psi.\label{eq: Ricci_scalar}
\end{eqnarray}
From this background, to obtain the gravitational field equation in
the conventional form, we combine (\ref{eq: Riccitensor1}) and (\ref{eq: Ricci_scalar})
\begin{eqnarray}
R_{ab}-\frac{1}{2}g_{ab}R & = & \frac{8\pi}{\alpha\left(z,q\right)\psi}T_{ab}+\frac{1}{\psi}\left(\nabla_{a}\nabla_{b}-g_{ab}\square\right)\psi.\label{eq:q-Einstein}
\end{eqnarray}
This is the field equation that corresponds to a kind of scalar-tensor
theory of gravity. 

At this stage, we consider the scalar field $\psi$ as a component
of the matter sector, contributing to the overall stress-energy content
of the system. Consequently, the total energy-momentum tensor $T_{ab}$
can be decomposed into two distinct parts:

\begin{equation}
T_{ab}=T_{ab}^{M}+\frac{\mathcal{\omega}}{8\pi\psi}\left(\partial_{a}\psi\partial_{b}\psi-\frac{1}{2}g_{ab}\partial_{c}\psi\partial^{c}\psi\right),
\end{equation}
where $T_{ab}^{M}$ originated from ordinary matter, the remaining
part is related to the scalar field and $\omega$ is an arbitrary
constant. Therefore, the field equation (\ref{eq:q-Einstein}) becomes
\begin{eqnarray}
R_{ab}-\frac{1}{2}g_{ab}R & = & \frac{8\pi}{\alpha\left(z,q\right)\psi}T_{ab}^{M}+\frac{\omega}{\alpha\left(z,q\right)\psi^{2}}\left(\partial_{a}\psi\partial_{b}\psi-\frac{1}{2}g_{ab}\partial_{c}\psi\partial^{c}\psi\right)+\frac{1}{\psi}\left(\nabla_{a}\nabla_{b}-g_{ab}\square\right)\psi.\label{eq: q-Einstein_BD}
\end{eqnarray}
where the $q$-deformed energy-momentum tensor for the BD scalar $T_{ab}^{\psi}$
can be defined by
\begin{equation}
T_{ab}^{\psi}=\frac{1}{8\pi}\left[\frac{\omega}{\alpha\left(z,q\right)\psi^{2}}\left(\partial_{a}\psi\partial_{b}\psi-\frac{1}{2}g_{ab}\partial_{c}\psi\partial^{c}\psi\right)+\frac{1}{\psi}\left(\nabla_{a}\nabla_{b}-g_{ab}\square\right)\psi\right].
\end{equation}

For consistency, in the case of $\alpha\left(z,q\right)=1$ the field
equation (\ref{eq: q-Einstein_BD}) reduces to Brans-Dicke-like theory
\citep{Brans:1961sx} or setting $\psi=1$ the equations go to $q$-deformed
Einstein field equations \citep{Senay:2018xaj,Kibaroglu:2018mnx,Kibaroglu:2019odt,SENAY2021136536}
and finally when we both set $\psi=1$ and $\alpha\left(z,q\right)=1$
the theory reduces to the well-known Einstein's theory of general
relativity together with matter. 

Furthermore, in our current formulation, the deformation function
$\alpha\left(q,z\right)$ is treated as a constant parameter that
encapsulates the effects of $q$-deformation and statistical considerations
from Verlinde\textquoteright s entropic gravity approach. As it appears
solely as a constant multiplicative factor, $\alpha\left(q,z\right)$
effectively rescales Newton's gravitational constant as $G_{eff}=G/\alpha\left(q,z\right)$,
without introducing additional dynamical degrees of freedom. Consequently,
the modified field equations retain the standard form of Brans-Dicke
equations with a constant scalar field, and the deformation does not
lead to any intrinsic modification of the gravitational dynamics.
In this context, diffeomorphism invariance and local Lorentz invariance
remain preserved. However, the constancy of $\alpha\left(q,z\right)$
implies that the theory does not generically recover the correct Newtonian
limit or pass solar system tests unless $\alpha\left(q,z\right)\simeq1$,
corresponding to the undeformed case $T=T_{U}$. 

Therefore, we have found a $q$-deformed scalar-tensor theory of gravity
by introducing the deformation function $\alpha\left(z,q\right)$
(see Eq.(\ref{eq:alpha})), which originates from the quantum algebra
in Eqs.(\ref{eq:q-algebra1}) and (\ref{eq:q-algebra2}). This framework
allows us to study scenarios particularly associated with quantum
effects or alternative theories of gravity.

\section{Conclusion\label{sec:Conclusion}}

In this study, we have developed a scalar-tensor theory of gravitation
by integrating Verlinde's entropic gravity approach, thereby extending
the conventional framework of gravitational theories. Additionally,
we have thoroughly examined the incorporation of $q$-deformed statistics
into this model, demonstrating how it provides a more generalized
and adaptable structure for analysing gravitational interactions.

In this scope, we have found the $q$-deformed field equation for
this scale-invariant model in (\ref{eq:q-Einstein}). This equation
may also be interpreted as the $q$-deformed BD gravitational field
equation by assuming that the introduced coupling constant $\omega$
corresponds to the BD parameter (\ref{eq: q-Einstein_BD}). We demonstrated
that the theory consistently recovers known frameworks in specific
limits: it reduces to the standard Brans-Dicke theory when the deformation
function is unity, and to the $q$-deformed Einstein gravity when
the scalar field is constant. In the fully undeformed case (when both
the scalar field is constant and the deformation function equals one)
the theory reverts to general relativity with matter. 

Importantly, the deformation function $\alpha\left(q,z\right)$ is
treated as a constant that encodes the effects of quantum deformation
and statistical considerations. As such, it serves to rescale the
effective gravitational constant without introducing additional dynamical
degrees of freedom. This formulation preserves both diffeomorphism
invariance and local Lorentz symmetry. However, we emphasize that
unless the deformation parameter is sufficiently close to unity, the
theory may not reproduce the correct Newtonian limit or pass solar
system constraints. Thus, the theory does not yield a substantial
modification to the standard Brans-Dicke theory and its potential
cosmological solutions. To address this limitation, a more refined
formulation may involve promoting the deformation function $\alpha\left(q,z\right)$
to a spacetime-dependent quantity; potentially through coupling with
a dynamical scalar field or thermodynamically motivated variables.
This extension could enhance the model\textquoteright s capacity to
recover the correct Newtonian limit and achieve better consistency
with observational constraints, an avenue we intend to investigate
in future work. Furthermore, beyond the Brans-Dicke framework, exploring
cosmological solutions derived from the modified field equations (\ref{eq:q-Einstein})
may yield novel insights into early universe dynamics and cosmic evolution,
which will be the subject of ongoing research.

\section*{Conflicts of Interest }

The authors declare no conflicts of interest.

\section*{Data Availability Statement }

No new data were created or analysed in this study. 

\section*{Declaration of generative AI and AI-assisted technologies in the
writing process}

During the preparation of this work the author used ChatGPT - GPT-4o
in order to improve readability of the text. After using this tool/service,
the author reviewed and edited the content as needed and takes full
responsibility for the content of the publication.

\section*{Appendix I: Derivation of $q$-deformed temperature\label{sec:Appendix:-Derivation-of-Temp}}

In this appendix, we show the derivation of $q$-deformed temperature.
The total entropy $\mathcal{S}$ remains unchanged when the entropic
force matches the external force that drives an increase in entropy.
Under these conditions, the system reaches statistical equilibrium,
where the entropy change becomes zero. This condition can be expressed
mathematically as 
\begin{equation}
\frac{\text{d}}{\text{d}x^{a}}S(E,x^{a})=0,\label{eq:sentropy1-1}
\end{equation}
which results in the following expression 
\begin{equation}
\frac{\partial S}{\partial E}\frac{\partial E}{\partial x^{a}}+\frac{\partial S}{\partial x^{a}}=0.\label{eq:sentropy2}
\end{equation}
where $\frac{\partial E}{\partial x^{a}}=-F_{a}$ and $\frac{\partial S}{\partial x^{a}}=\nabla_{a}S$.
Using the relation for the change in entropy (\ref{eq:q-entropy1})
\begin{equation}
\frac{dS}{dE}=\frac{5(2\pi m)^{3/2}VE^{3/2}}{2Th^{3}}\left[\frac{5}{2}f_{5/2}(z,q)-f_{3/2}(z,q)\ln z\right],
\end{equation}
we arrive at the equation:

\begin{equation}
\frac{5(2\pi m)^{3/2}VE^{3/2}}{2h^{3}}\left[\frac{5}{2}f_{5/2}(z,q)-f_{3/2}(z,q)\ln z\right]F_{a}=T\nabla_{a}S.\label{eq:equality1}
\end{equation}
Substituting $F_{a}=-me^{\phi}\nabla_{a}\phi$ and $\nabla_{a}S=-\frac{2\pi mN_{a}}{\hbar}$
into this equation, we obtain: 
\begin{equation}
\frac{5(2\pi m)^{3/2}VE^{3/2}}{2h^{3}}\left[\frac{5}{2}f_{5/2}(z,q)-f_{3/2}(z,q)\ln z\right]me^{\phi}\nabla_{a}\phi=T\frac{4\pi^{2}mN_{a}}{h}.\label{eq:equality2}
\end{equation}
Rearranging this expression, we obtain:

\begin{equation}
T=\frac{5(2m\pi E)^{3/2}V\left[\frac{5}{2}f_{5/2}(z,q)-f_{3/2}(z,q)\ln z\right]}{2h^{3}}\frac{\hbar N^{a}e^{\phi}\nabla_{a}\phi}{2\pi},\label{eq:equality3}
\end{equation}
where we have used $N_{a}N^{a}=1$. Therefore, we define the $q$-deformed
temperature as: 
\begin{equation}
T=\alpha\left(z,q\right)\frac{\hbar}{2\pi}e^{\phi}N^{a}\nabla_{a}\phi,\label{eq:equality3}
\end{equation}
where 
\begin{equation}
\alpha\left(z,q\right)=\frac{5(2m\pi E)^{3/2}V\left[\frac{5}{2}f_{5/2}(z,q)-f_{3/2}(z,q)\ln z\right]}{2h^{3}}.
\end{equation}
Finally, using $E=T$, $\lambda=h/(2\pi mT)^{1/2}$ and $\frac{Vf_{3/2}(z,q)}{\lambda^{3}}=N$,
the function $\alpha(z,q)$ can be written as 
\begin{equation}
\frac{\alpha\left(z,q\right)}{N}=\frac{5}{2}\left[\frac{5}{2}\frac{f_{5/2}(z,q)}{f_{3/2}(z,q)}-\ln z\right].
\end{equation}

\section*{Appendix II: Derivation of Eq. (\ref{eq:q-mass1})}

Let us analyze the derivation of Eq.(\ref{eq:q-mass}) in detail.
We begin with the following expression for the mass:

\begin{equation}
M=\frac{\alpha\left(z,q\right)}{4\pi}\oint_{\partial\Sigma}e^{\phi}N^{a}\left(a_{a}+\frac{1}{2}\nabla_{a}\log\psi\right)\psi\text{d}A,
\end{equation}
Using the relations $a^{a}=e^{-2\phi}\xi^{b}\nabla_{b}\xi^{a}$ and
$\nabla_{a}\log\psi=\frac{1}{\psi}\nabla_{a}\psi$ , we rewrite the
integrand as:

\begin{eqnarray}
 & = & \frac{\alpha\left(z,q\right)}{4\pi}\oint_{\partial\Sigma}N^{a}\left(\psi\frac{\xi^{b}}{e^{\phi}}\nabla_{b}\xi_{a}+\frac{1}{2}e^{\phi}\nabla_{a}\psi\right)\text{d}A.
\end{eqnarray}
Introducing the unit time-like normal vector $n^{a}=\frac{\xi^{b}}{e^{\phi}}=\frac{\xi^{b}}{\sqrt{-\xi^{a}\xi_{a}}}$
yielding:

\begin{eqnarray}
 & = & \frac{\alpha\left(z,q\right)}{4\pi}\oint_{\partial\Sigma}\left(\psi N^{a}n^{b}\nabla_{b}\xi_{a}-\frac{1}{2}\xi_{b}n^{b}N^{a}\nabla_{a}\psi\right)\text{d}A,
\end{eqnarray}
This can be expressed in antisymmetric form as:

\begin{eqnarray}
 & = & \frac{\alpha\left(z,q\right)}{4\pi}\oint_{\partial\Sigma}\left(\psi\nabla^{[a}\xi^{b]}-\frac{1}{2}\xi^{[a}\nabla^{b]}\psi\right)\text{d}A_{ab},\label{eq: Mass1-1-1}
\end{eqnarray}
where $dA_{ab}=N_{ab}dA$ is an anti-symmetric tensor that represents
the surface element on the boundary $\partial\Sigma$ and $N_{ab}=n_{[a}N_{b]}$
is the normal \textquotedbl bi-vector\textquotedbl{} to $\Sigma$.
Applying Stokes\textquoteright{} theorem, we convert the surface integral
into a volume integral:

\begin{eqnarray}
M & = & \frac{\alpha\left(z,q\right)}{4\pi}\int_{\Sigma}\nabla_{b}\left(\psi\nabla^{[a}\xi^{b]}-\frac{1}{2}\xi^{[a}\nabla^{b]}\psi\right)\text{d}\Sigma_{a}.
\end{eqnarray}
We now consider each term in the integrand separately. For the first
term:

\begin{eqnarray}
\frac{\alpha\left(z,q\right)}{4\pi}\int_{\Sigma}\nabla_{b}\left(\psi\nabla^{[a}\xi^{b]}\right)\text{d}\Sigma_{a} & = & \frac{\alpha\left(z,q\right)}{4\pi}\oint_{\Sigma}\left(\nabla_{b}\psi\nabla^{[a}\xi^{b]}+\psi\nabla_{b}\nabla^{[a}\xi^{b]}\right)\text{d}\Sigma_{a},\nonumber \\
 & = & \frac{\alpha\left(z,q\right)}{4\pi}\oint_{\Sigma}\left(\nabla_{b}\psi\nabla^{a}\xi^{b}-\psi\nabla_{b}\nabla^{b}\xi^{a}\right)\text{d}\Sigma_{a},
\end{eqnarray}
using $R_{\,\,b}^{a}\xi^{b}=-\nabla_{b}\nabla^{b}\xi^{a}$ and the
identity $\nabla_{b}\psi\nabla^{a}\xi^{b}=\nabla^{a}\left(\nabla_{b}\psi\xi^{b}\right)-\left(\nabla^{a}\nabla_{b}\psi\right)\xi^{b}$,
we obtain;

\begin{eqnarray}
 & = & \frac{\alpha\left(z,q\right)}{4\pi}\int_{\Sigma}\left[\nabla^{a}\left(\nabla_{b}\psi\xi^{b}\right)-\nabla^{a}\nabla_{b}\psi\xi^{b}+\psi R_{\,\,b}^{a}\xi^{b}\right]\text{d}\Sigma_{a}.
\end{eqnarray}
Neglecting total derivative terms, we arrive at:

\begin{eqnarray}
 & = & \frac{\alpha\left(z,q\right)}{4\pi}\int_{\Sigma}\left[-\left(\nabla^{a}\nabla_{b}\psi\right)\xi^{b}+\psi R_{\,\,b}^{a}\xi^{b}\right]\text{d}\Sigma_{a}.\label{eq: Mass2_1}
\end{eqnarray}
Now, for the second term:

\begin{eqnarray}
-\frac{\alpha\left(z,q\right)}{8\pi}\int_{\Sigma}\nabla_{b}\left(\xi^{[a}\nabla^{b]}\psi\right)\text{d}\Sigma_{a} & = & -\frac{\alpha\left(z,q\right)}{8\pi}\int_{\Sigma}\left(\nabla_{b}\xi^{a}\nabla^{b}\psi+\xi^{a}\square\psi-\xi^{b}\nabla_{b}\nabla^{a}\psi\right)\text{d}\Sigma_{a},
\end{eqnarray}
where we used $\nabla_{b}\xi^{b}=0$ and $\square=g^{ab}\nabla_{a}\nabla_{b}$
denotes the d'Alembertian operator. Now, using the identity $\xi^{b}\nabla_{b}\nabla^{a}\psi=\nabla^{a}\left(\xi^{b}\nabla_{b}\psi\right)-\nabla^{a}\xi^{b}\nabla_{b}\psi$,
we find:

\begin{eqnarray}
 & = & -\frac{\alpha\left(z,q\right)}{8\pi}\int_{\Sigma}\left(\nabla_{b}\xi^{a}\nabla^{b}\psi+\xi^{a}\square\psi-\nabla^{a}\left(\xi^{b}\nabla_{b}\psi\right)+\nabla^{a}\xi^{b}\nabla_{b}\psi\right)\text{d}\Sigma_{a},
\end{eqnarray}
using the Killing equation $\nabla_{a}\xi_{b}=-\nabla_{b}\xi_{a}$
and simplifying, we obtain;

\begin{eqnarray}
 & = & -\frac{\alpha\left(z,q\right)}{8\pi}\int_{\Sigma}g_{ab}\square\psi\xi^{b}\text{d}\Sigma^{a}.\label{eq: Mass2_2}
\end{eqnarray}
Finally, combining both contributions (\ref{eq: Mass2_1}) and (\ref{eq: Mass2_2}),
we arrive at the final expression:
\begin{equation}
M=\frac{\alpha\left(z,q\right)}{4\pi G}\int_{\Sigma}\left[\psi R_{ab}-\left(\nabla_{a}\nabla_{b}+\frac{1}{2}g_{ab}\square\right)\psi\right]\xi^{b}\text{d}\Sigma^{a},\label{eq:q-mass2}
\end{equation}
which corresponds to Eq.(\ref{eq:q-mass1}).

\bibliography{Q_BD}

\begin{thebibliography}{78}%
\makeatletter
\providecommand \@ifxundefined [1]{%
 \@ifx{#1\undefined}
}%
\providecommand \@ifnum [1]{%
 \ifnum #1\expandafter \@firstoftwo
 \else \expandafter \@secondoftwo
 \fi
}%
\providecommand \@ifx [1]{%
 \ifx #1\expandafter \@firstoftwo
 \else \expandafter \@secondoftwo
 \fi
}%
\providecommand \natexlab [1]{#1}%
\providecommand \enquote  [1]{``#1''}%
\providecommand \bibnamefont  [1]{#1}%
\providecommand \bibfnamefont [1]{#1}%
\providecommand \citenamefont [1]{#1}%
\providecommand \href@noop [0]{\@secondoftwo}%
\providecommand \href [0]{\begingroup \@sanitize@url \@href}%
\providecommand \@href[1]{\@@startlink{#1}\@@href}%
\providecommand \@@href[1]{\endgroup#1\@@endlink}%
\providecommand \@sanitize@url [0]{\catcode `\\12\catcode `\$12\catcode
  `\&12\catcode `\#12\catcode `\^12\catcode `\_12\catcode `\%12\relax}%
\providecommand \@@startlink[1]{}%
\providecommand \@@endlink[0]{}%
\providecommand \url  [0]{\begingroup\@sanitize@url \@url }%
\providecommand \@url [1]{\endgroup\@href {#1}{\urlprefix }}%
\providecommand \urlprefix  [0]{URL }%
\providecommand \Eprint [0]{\href }%
\providecommand \doibase [0]{https://doi.org/}%
\providecommand \selectlanguage [0]{\@gobble}%
\providecommand \bibinfo  [0]{\@secondoftwo}%
\providecommand \bibfield  [0]{\@secondoftwo}%
\providecommand \translation [1]{[#1]}%
\providecommand \BibitemOpen [0]{}%
\providecommand \bibitemStop [0]{}%
\providecommand \bibitemNoStop [0]{.\EOS\space}%
\providecommand \EOS [0]{\spacefactor3000\relax}%
\providecommand \BibitemShut  [1]{\csname bibitem#1\endcsname}%
\let\auto@bib@innerbib\@empty
\bibitem [{\citenamefont {Capozziello}\ and\ \citenamefont
  {De~Laurentis}(2011)}]{Capozziello:2011Extended}%
  \BibitemOpen
  \bibfield  {author} {\bibinfo {author} {\bibfnamefont {S.}~\bibnamefont
  {Capozziello}}\ and\ \bibinfo {author} {\bibfnamefont {M.}~\bibnamefont
  {De~Laurentis}},\ }\bibfield  {title} {\bibinfo {title} {{Extended Theories
  of Gravity}},\ }\href {https://doi.org/10.1016/j.physrep.2011.09.003}
  {\bibfield  {journal} {\bibinfo  {journal} {Phys. Rept.}\ }\textbf {\bibinfo
  {volume} {509}},\ \bibinfo {pages} {167} (\bibinfo {year}
  {2011})}\BibitemShut {NoStop}%
\bibitem [{\citenamefont {Clifton}\ \emph {et~al.}(2012)\citenamefont
  {Clifton}, \citenamefont {Ferreira}, \citenamefont {Padilla},\ and\
  \citenamefont {Skordis}}]{Clifton:2011Modified}%
  \BibitemOpen
  \bibfield  {author} {\bibinfo {author} {\bibfnamefont {T.}~\bibnamefont
  {Clifton}}, \bibinfo {author} {\bibfnamefont {P.~G.}\ \bibnamefont
  {Ferreira}}, \bibinfo {author} {\bibfnamefont {A.}~\bibnamefont {Padilla}},\
  and\ \bibinfo {author} {\bibfnamefont {C.}~\bibnamefont {Skordis}},\
  }\bibfield  {title} {\bibinfo {title} {{Modified Gravity and Cosmology}},\
  }\href {https://doi.org/10.1016/j.physrep.2012.01.001} {\bibfield  {journal}
  {\bibinfo  {journal} {Phys. Rept.}\ }\textbf {\bibinfo {volume} {513}},\
  \bibinfo {pages} {1} (\bibinfo {year} {2012})}\BibitemShut {NoStop}%
\bibitem [{\citenamefont {Nojiri}\ \emph {et~al.}(2017)\citenamefont {Nojiri},
  \citenamefont {Odintsov},\ and\ \citenamefont
  {Oikonomou}}]{Nojiri:2017Modified}%
  \BibitemOpen
  \bibfield  {author} {\bibinfo {author} {\bibfnamefont {S.}~\bibnamefont
  {Nojiri}}, \bibinfo {author} {\bibfnamefont {S.~D.}\ \bibnamefont
  {Odintsov}},\ and\ \bibinfo {author} {\bibfnamefont {V.~K.}\ \bibnamefont
  {Oikonomou}},\ }\bibfield  {title} {\bibinfo {title} {{Modified Gravity
  Theories on a Nutshell: Inflation, Bounce and Late-time Evolution}},\ }\href
  {https://doi.org/10.1016/j.physrep.2017.06.001} {\bibfield  {journal}
  {\bibinfo  {journal} {Phys. Rept.}\ }\textbf {\bibinfo {volume} {692}},\
  \bibinfo {pages} {1} (\bibinfo {year} {2017})}\BibitemShut {NoStop}%
\bibitem [{\citenamefont {Dirac}(1936)}]{Dirac:1936Wave}%
  \BibitemOpen
  \bibfield  {author} {\bibinfo {author} {\bibfnamefont {P.~A.~M.}\
  \bibnamefont {Dirac}},\ }\bibfield  {title} {\bibinfo {title} {{Wave
  equations in conformal space}},\ }\href {https://doi.org/10.2307/1968455}
  {\bibfield  {journal} {\bibinfo  {journal} {Annals Math.}\ }\textbf {\bibinfo
  {volume} {37}},\ \bibinfo {pages} {429} (\bibinfo {year} {1936})}\BibitemShut
  {NoStop}%
\bibitem [{\citenamefont {Mack}\ and\ \citenamefont
  {Salam}(1969)}]{Mack:1969Finite}%
  \BibitemOpen
  \bibfield  {author} {\bibinfo {author} {\bibfnamefont {G.}~\bibnamefont
  {Mack}}\ and\ \bibinfo {author} {\bibfnamefont {A.}~\bibnamefont {Salam}},\
  }\bibfield  {title} {\bibinfo {title} {{Finite component field
  representations of the conformal group}},\ }\href
  {https://doi.org/10.1016/0003-4916(69)90278-4} {\bibfield  {journal}
  {\bibinfo  {journal} {Annals Phys.}\ }\textbf {\bibinfo {volume} {53}},\
  \bibinfo {pages} {174} (\bibinfo {year} {1969})}\BibitemShut {NoStop}%
\bibitem [{\citenamefont {Shaposhnikov}\ and\ \citenamefont
  {Zenhausern}(2009)}]{Shaposhnikov:2009Scale}%
  \BibitemOpen
  \bibfield  {author} {\bibinfo {author} {\bibfnamefont {M.}~\bibnamefont
  {Shaposhnikov}}\ and\ \bibinfo {author} {\bibfnamefont {D.}~\bibnamefont
  {Zenhausern}},\ }\bibfield  {title} {\bibinfo {title} {{Scale invariance,
  unimodular gravity and dark energy}},\ }\href
  {https://doi.org/10.1016/j.physletb.2008.11.054} {\bibfield  {journal}
  {\bibinfo  {journal} {Phys. Lett. B}\ }\textbf {\bibinfo {volume} {671}},\
  \bibinfo {pages} {187} (\bibinfo {year} {2009})},\ \Eprint
  {https://arxiv.org/abs/0809.3395} {arXiv:0809.3395 [hep-th]} \BibitemShut
  {NoStop}%
\bibitem [{\citenamefont {Israelit}(2011)}]{Israelit:2011Weyl}%
  \BibitemOpen
  \bibfield  {author} {\bibinfo {author} {\bibfnamefont {M.}~\bibnamefont
  {Israelit}},\ }\bibfield  {title} {\bibinfo {title} {{A Weyl-Dirac
  Cosmological Model with DM and DE}},\ }\href
  {https://doi.org/10.1007/s10714-010-1092-3} {\bibfield  {journal} {\bibinfo
  {journal} {Gen. Rel. Grav.}\ }\textbf {\bibinfo {volume} {43}},\ \bibinfo
  {pages} {751} (\bibinfo {year} {2011})}\BibitemShut {NoStop}%
\bibitem [{\citenamefont {Bars}\ \emph {et~al.}(2014)\citenamefont {Bars},
  \citenamefont {Steinhardt},\ and\ \citenamefont
  {Turok}}]{Bars:2013LocalConf}%
  \BibitemOpen
  \bibfield  {author} {\bibinfo {author} {\bibfnamefont {I.}~\bibnamefont
  {Bars}}, \bibinfo {author} {\bibfnamefont {P.}~\bibnamefont {Steinhardt}},\
  and\ \bibinfo {author} {\bibfnamefont {N.}~\bibnamefont {Turok}},\ }\bibfield
   {title} {\bibinfo {title} {{Local Conformal Symmetry in Physics and
  Cosmology}},\ }\href {https://doi.org/10.1103/PhysRevD.89.043515} {\bibfield
  {journal} {\bibinfo  {journal} {Phys. Rev. D}\ }\textbf {\bibinfo {volume}
  {89}},\ \bibinfo {pages} {043515} (\bibinfo {year} {2014})}\BibitemShut
  {NoStop}%
\bibitem [{\citenamefont {Aguila}\ \emph {et~al.}(2014)\citenamefont {Aguila},
  \citenamefont {Madriz~Aguilar}, \citenamefont {Moreno},\ and\ \citenamefont
  {Bellini}}]{Aguila:2014PresentAccelerated}%
  \BibitemOpen
  \bibfield  {author} {\bibinfo {author} {\bibfnamefont {R.}~\bibnamefont
  {Aguila}}, \bibinfo {author} {\bibfnamefont {J.~E.}\ \bibnamefont
  {Madriz~Aguilar}}, \bibinfo {author} {\bibfnamefont {C.}~\bibnamefont
  {Moreno}},\ and\ \bibinfo {author} {\bibfnamefont {M.}~\bibnamefont
  {Bellini}},\ }\bibfield  {title} {\bibinfo {title} {{Present accelerated
  expansion of the universe from new Weyl-Integrable gravity approach}},\
  }\href {https://doi.org/10.1140/epjc/s10052-014-3158-y} {\bibfield  {journal}
  {\bibinfo  {journal} {Eur. Phys. J. C}\ }\textbf {\bibinfo {volume} {74}},\
  \bibinfo {pages} {3158} (\bibinfo {year} {2014})}\BibitemShut {NoStop}%
\bibitem [{\citenamefont {Karananas}\ and\ \citenamefont
  {Shaposhnikov}(2016)}]{Karananas:2016Scale}%
  \BibitemOpen
  \bibfield  {author} {\bibinfo {author} {\bibfnamefont {G.~K.}\ \bibnamefont
  {Karananas}}\ and\ \bibinfo {author} {\bibfnamefont {M.}~\bibnamefont
  {Shaposhnikov}},\ }\bibfield  {title} {\bibinfo {title} {{Scale invariant
  alternatives to general relativity. II. Dilaton properties}},\ }\href
  {https://doi.org/10.1103/PhysRevD.93.084052} {\bibfield  {journal} {\bibinfo
  {journal} {Phys. Rev. D}\ }\textbf {\bibinfo {volume} {93}},\ \bibinfo
  {pages} {084052} (\bibinfo {year} {2016})}\BibitemShut {NoStop}%
\bibitem [{\citenamefont {Ferreira}\ \emph {et~al.}(2018)\citenamefont
  {Ferreira}, \citenamefont {Hill}, \citenamefont {Noller},\ and\ \citenamefont
  {Ross}}]{Ferreira:2018Inflation}%
  \BibitemOpen
  \bibfield  {author} {\bibinfo {author} {\bibfnamefont {P.~G.}\ \bibnamefont
  {Ferreira}}, \bibinfo {author} {\bibfnamefont {C.~T.}\ \bibnamefont {Hill}},
  \bibinfo {author} {\bibfnamefont {J.}~\bibnamefont {Noller}},\ and\ \bibinfo
  {author} {\bibfnamefont {G.~G.}\ \bibnamefont {Ross}},\ }\bibfield  {title}
  {\bibinfo {title} {{Inflation in a scale invariant universe}},\ }\href
  {https://doi.org/10.1103/PhysRevD.97.123516} {\bibfield  {journal} {\bibinfo
  {journal} {Phys. Rev. D}\ }\textbf {\bibinfo {volume} {97}},\ \bibinfo
  {pages} {123516} (\bibinfo {year} {2018})}\BibitemShut {NoStop}%
\bibitem [{\citenamefont {Casas}\ \emph {et~al.}(2019)\citenamefont {Casas},
  \citenamefont {Karananas}, \citenamefont {Pauly},\ and\ \citenamefont
  {Rubio}}]{Casas:2019Scale}%
  \BibitemOpen
  \bibfield  {author} {\bibinfo {author} {\bibfnamefont {S.}~\bibnamefont
  {Casas}}, \bibinfo {author} {\bibfnamefont {G.~K.}\ \bibnamefont
  {Karananas}}, \bibinfo {author} {\bibfnamefont {M.}~\bibnamefont {Pauly}},\
  and\ \bibinfo {author} {\bibfnamefont {J.}~\bibnamefont {Rubio}},\ }\bibfield
   {title} {\bibinfo {title} {{Scale-invariant alternatives to general
  relativity. III. The inflation-dark energy connection}},\ }\href
  {https://doi.org/10.1103/PhysRevD.99.063512} {\bibfield  {journal} {\bibinfo
  {journal} {Phys. Rev. D}\ }\textbf {\bibinfo {volume} {99}},\ \bibinfo
  {pages} {063512} (\bibinfo {year} {2019})}\BibitemShut {NoStop}%
\bibitem [{\citenamefont {Ghilencea}(2021)}]{Ghilencea:2021GaugingScale}%
  \BibitemOpen
  \bibfield  {author} {\bibinfo {author} {\bibfnamefont {D.~M.}\ \bibnamefont
  {Ghilencea}},\ }\bibfield  {title} {\bibinfo {title} {{Gauging scale symmetry
  and inflation: Weyl versus Palatini gravity}},\ }\href
  {https://doi.org/10.1140/epjc/s10052-021-09226-1} {\bibfield  {journal}
  {\bibinfo  {journal} {Eur. Phys. J. C}\ }\textbf {\bibinfo {volume} {81}},\
  \bibinfo {pages} {510} (\bibinfo {year} {2021})}\BibitemShut {NoStop}%
\bibitem [{\citenamefont {Tang}\ and\ \citenamefont
  {Wu}(2020)}]{Tang:2020WeylSymmetry}%
  \BibitemOpen
  \bibfield  {author} {\bibinfo {author} {\bibfnamefont {Y.}~\bibnamefont
  {Tang}}\ and\ \bibinfo {author} {\bibfnamefont {Y.-L.}\ \bibnamefont {Wu}},\
  }\bibfield  {title} {\bibinfo {title} {{Weyl Symmetry Inspired Inflation and
  Dark Matter}},\ }\href {https://doi.org/10.1016/j.physletb.2020.135320}
  {\bibfield  {journal} {\bibinfo  {journal} {Phys. Lett. B}\ }\textbf
  {\bibinfo {volume} {803}},\ \bibinfo {pages} {135320} (\bibinfo {year}
  {2020})}\BibitemShut {NoStop}%
\bibitem [{\citenamefont {Bekenstein}(1973)}]{Bekenstein1973}%
  \BibitemOpen
  \bibfield  {author} {\bibinfo {author} {\bibfnamefont {J.~D.}\ \bibnamefont
  {Bekenstein}},\ }\bibfield  {title} {\bibinfo {title} {Black holes and
  entropy},\ }\href {https://doi.org/10.1103/PhysRevD.7.2333} {\bibfield
  {journal} {\bibinfo  {journal} {Phys. Rev. D}\ }\textbf {\bibinfo {volume}
  {7}},\ \bibinfo {pages} {2333} (\bibinfo {year} {1973})}\BibitemShut
  {NoStop}%
\bibitem [{\citenamefont {Hawking}(1975)}]{Hawking:1975vcx}%
  \BibitemOpen
  \bibfield  {author} {\bibinfo {author} {\bibfnamefont {S.~W.}\ \bibnamefont
  {Hawking}},\ }\bibfield  {title} {\bibinfo {title} {{Particle Creation by
  Black Holes}},\ }\href {https://doi.org/10.1007/BF02345020} {\bibfield
  {journal} {\bibinfo  {journal} {Commun. Math. Phys.}\ }\textbf {\bibinfo
  {volume} {43}},\ \bibinfo {pages} {199} (\bibinfo {year} {1975})},\ \bibinfo
  {note} {[Erratum: Commun.Math.Phys. 46, 206 (1976)]}\BibitemShut {NoStop}%
\bibitem [{\citenamefont {Jacobson}(1995)}]{Jacobson:1995Thermodynamics}%
  \BibitemOpen
  \bibfield  {author} {\bibinfo {author} {\bibfnamefont {T.}~\bibnamefont
  {Jacobson}},\ }\bibfield  {title} {\bibinfo {title} {{Thermodynamics of
  space-time: The Einstein equation of state}},\ }\href
  {https://doi.org/10.1103/PhysRevLett.75.1260} {\bibfield  {journal} {\bibinfo
   {journal} {Phys. Rev. Lett.}\ }\textbf {\bibinfo {volume} {75}},\ \bibinfo
  {pages} {1260} (\bibinfo {year} {1995})}\BibitemShut {NoStop}%
\bibitem [{\citenamefont {Padmanabhan}(2005)}]{Padmanabhan:2005Gravity}%
  \BibitemOpen
  \bibfield  {author} {\bibinfo {author} {\bibfnamefont {T.}~\bibnamefont
  {Padmanabhan}},\ }\bibfield  {title} {\bibinfo {title} {{Gravity and the
  thermodynamics of horizons}},\ }\href
  {https://doi.org/10.1016/j.physrep.2004.10.003} {\bibfield  {journal}
  {\bibinfo  {journal} {Phys. Rept.}\ }\textbf {\bibinfo {volume} {406}},\
  \bibinfo {pages} {49} (\bibinfo {year} {2005})}\BibitemShut {NoStop}%
\bibitem [{\citenamefont
  {Padmanabhan}(2010)}]{Padmanabhan:2010Thermodynamical}%
  \BibitemOpen
  \bibfield  {author} {\bibinfo {author} {\bibfnamefont {T.}~\bibnamefont
  {Padmanabhan}},\ }\bibfield  {title} {\bibinfo {title} {{Thermodynamical
  Aspects of Gravity: New insights}},\ }\href
  {https://doi.org/10.1088/0034-4885/73/4/046901} {\bibfield  {journal}
  {\bibinfo  {journal} {Rept. Prog. Phys.}\ }\textbf {\bibinfo {volume} {73}},\
  \bibinfo {pages} {046901} (\bibinfo {year} {2010})}\BibitemShut {NoStop}%
\bibitem [{\citenamefont {Cai}\ and\ \citenamefont
  {Kim}(2005)}]{Cai:2005FirstLaw}%
  \BibitemOpen
  \bibfield  {author} {\bibinfo {author} {\bibfnamefont {R.-G.}\ \bibnamefont
  {Cai}}\ and\ \bibinfo {author} {\bibfnamefont {S.~P.}\ \bibnamefont {Kim}},\
  }\bibfield  {title} {\bibinfo {title} {{First law of thermodynamics and
  Friedmann equations of Friedmann-Robertson-Walker universe}},\ }\href
  {https://doi.org/10.1088/1126-6708/2005/02/050} {\bibfield  {journal}
  {\bibinfo  {journal} {JHEP}\ }\textbf {\bibinfo {volume} {02}},\ \bibinfo
  {pages} {050}}\BibitemShut {NoStop}%
\bibitem [{\citenamefont {Akbar}\ and\ \citenamefont
  {Cai}(2007)}]{Akbar:2007Thermodynamic}%
  \BibitemOpen
  \bibfield  {author} {\bibinfo {author} {\bibfnamefont {M.}~\bibnamefont
  {Akbar}}\ and\ \bibinfo {author} {\bibfnamefont {R.-G.}\ \bibnamefont
  {Cai}},\ }\bibfield  {title} {\bibinfo {title} {{Thermodynamic Behavior of
  Friedmann Equations at Apparent Horizon of FRW Universe}},\ }\href
  {https://doi.org/10.1103/PhysRevD.75.084003} {\bibfield  {journal} {\bibinfo
  {journal} {Phys. Rev. D}\ }\textbf {\bibinfo {volume} {75}},\ \bibinfo
  {pages} {084003} (\bibinfo {year} {2007})}\BibitemShut {NoStop}%
\bibitem [{\citenamefont {Cai}\ and\ \citenamefont
  {Cao}(2007)}]{Cai:2007Unified}%
  \BibitemOpen
  \bibfield  {author} {\bibinfo {author} {\bibfnamefont {R.-G.}\ \bibnamefont
  {Cai}}\ and\ \bibinfo {author} {\bibfnamefont {L.-M.}\ \bibnamefont {Cao}},\
  }\bibfield  {title} {\bibinfo {title} {{Unified first law and thermodynamics
  of apparent horizon in FRW universe}},\ }\href
  {https://doi.org/10.1103/PhysRevD.75.064008} {\bibfield  {journal} {\bibinfo
  {journal} {Phys. Rev. D}\ }\textbf {\bibinfo {volume} {75}},\ \bibinfo
  {pages} {064008} (\bibinfo {year} {2007})}\BibitemShut {NoStop}%
\bibitem [{\citenamefont {Jamil}\ \emph {et~al.}(2010)\citenamefont {Jamil},
  \citenamefont {Saridakis},\ and\ \citenamefont
  {Setare}}]{Jamil:2010Thermodynamics}%
  \BibitemOpen
  \bibfield  {author} {\bibinfo {author} {\bibfnamefont {M.}~\bibnamefont
  {Jamil}}, \bibinfo {author} {\bibfnamefont {E.~N.}\ \bibnamefont
  {Saridakis}},\ and\ \bibinfo {author} {\bibfnamefont {M.~R.}\ \bibnamefont
  {Setare}},\ }\bibfield  {title} {\bibinfo {title} {{Thermodynamics of dark
  energy interacting with dark matter and radiation}},\ }\href
  {https://doi.org/10.1103/PhysRevD.81.023007} {\bibfield  {journal} {\bibinfo
  {journal} {Phys. Rev. D}\ }\textbf {\bibinfo {volume} {81}},\ \bibinfo
  {pages} {023007} (\bibinfo {year} {2010})}\BibitemShut {NoStop}%
\bibitem [{\citenamefont
  {Sheykhi}(2010{\natexlab{a}})}]{Sheykhi:2010Thermodynamics}%
  \BibitemOpen
  \bibfield  {author} {\bibinfo {author} {\bibfnamefont {A.}~\bibnamefont
  {Sheykhi}},\ }\bibfield  {title} {\bibinfo {title} {{Thermodynamics of
  apparent horizon and modified Friedmann equations}},\ }\href
  {https://doi.org/10.1140/epjc/s10052-010-1372-9} {\bibfield  {journal}
  {\bibinfo  {journal} {Eur. Phys. J. C}\ }\textbf {\bibinfo {volume} {69}},\
  \bibinfo {pages} {265} (\bibinfo {year} {2010}{\natexlab{a}})},\ \Eprint
  {https://arxiv.org/abs/1012.0383} {arXiv:1012.0383 [hep-th]} \BibitemShut
  {NoStop}%
\bibitem [{\citenamefont {Nojiri}\ \emph
  {et~al.}(2022{\natexlab{a}})\citenamefont {Nojiri}, \citenamefont
  {Odintsov},\ and\ \citenamefont {Paul}}]{Nojiri:2022Early}%
  \BibitemOpen
  \bibfield  {author} {\bibinfo {author} {\bibfnamefont {S.}~\bibnamefont
  {Nojiri}}, \bibinfo {author} {\bibfnamefont {S.~D.}\ \bibnamefont
  {Odintsov}},\ and\ \bibinfo {author} {\bibfnamefont {T.}~\bibnamefont
  {Paul}},\ }\bibfield  {title} {\bibinfo {title} {{Early and late universe
  holographic cosmology from a new generalized entropy}},\ }\href
  {https://doi.org/10.1016/j.physletb.2022.137189} {\bibfield  {journal}
  {\bibinfo  {journal} {Phys. Lett. B}\ }\textbf {\bibinfo {volume} {831}},\
  \bibinfo {pages} {137189} (\bibinfo {year} {2022}{\natexlab{a}})}\BibitemShut
  {NoStop}%
\bibitem [{\citenamefont {Nojiri}\ \emph
  {et~al.}(2022{\natexlab{b}})\citenamefont {Nojiri}, \citenamefont
  {Odintsov},\ and\ \citenamefont {Paul}}]{Nojiri:2022Modified}%
  \BibitemOpen
  \bibfield  {author} {\bibinfo {author} {\bibfnamefont {S.}~\bibnamefont
  {Nojiri}}, \bibinfo {author} {\bibfnamefont {S.~D.}\ \bibnamefont
  {Odintsov}},\ and\ \bibinfo {author} {\bibfnamefont {T.}~\bibnamefont
  {Paul}},\ }\bibfield  {title} {\bibinfo {title} {{Modified cosmology from the
  thermodynamics of apparent horizon}},\ }\href
  {https://doi.org/10.1016/j.physletb.2022.137553} {\bibfield  {journal}
  {\bibinfo  {journal} {Phys. Lett. B}\ }\textbf {\bibinfo {volume} {835}},\
  \bibinfo {pages} {137553} (\bibinfo {year} {2022}{\natexlab{b}})}\BibitemShut
  {NoStop}%
\bibitem [{\citenamefont {Nojiri}\ \emph {et~al.}(2024)\citenamefont {Nojiri},
  \citenamefont {Odintsov},\ and\ \citenamefont {Paul}}]{Nojiri:2024Different}%
  \BibitemOpen
  \bibfield  {author} {\bibinfo {author} {\bibfnamefont {S.}~\bibnamefont
  {Nojiri}}, \bibinfo {author} {\bibfnamefont {S.~D.}\ \bibnamefont
  {Odintsov}},\ and\ \bibinfo {author} {\bibfnamefont {T.}~\bibnamefont
  {Paul}},\ }\bibfield  {title} {\bibinfo {title} {{Different Aspects of
  Entropic Cosmology}},\ }\href {https://doi.org/10.3390/universe10090352}
  {\bibfield  {journal} {\bibinfo  {journal} {Universe}\ }\textbf {\bibinfo
  {volume} {10}},\ \bibinfo {pages} {352} (\bibinfo {year} {2024})}\BibitemShut
  {NoStop}%
\bibitem [{\citenamefont {Verlinde}(2011)}]{Verlinde:2010hp}%
  \BibitemOpen
  \bibfield  {author} {\bibinfo {author} {\bibfnamefont {E.~P.}\ \bibnamefont
  {Verlinde}},\ }\bibfield  {title} {\bibinfo {title} {{On the Origin of
  Gravity and the Laws of Newton}},\ }\href
  {https://doi.org/10.1007/JHEP04(2011)029} {\bibfield  {journal} {\bibinfo
  {journal} {JHEP}\ }\textbf {\bibinfo {volume} {04}},\ \bibinfo {pages}
  {029}}\BibitemShut {NoStop}%
\bibitem [{\citenamefont {Verlinde}(2017)}]{Verlinde:2016toy}%
  \BibitemOpen
  \bibfield  {author} {\bibinfo {author} {\bibfnamefont {E.~P.}\ \bibnamefont
  {Verlinde}},\ }\bibfield  {title} {\bibinfo {title} {{Emergent Gravity and
  the Dark Universe}},\ }\href {https://doi.org/10.21468/SciPostPhys.2.3.016}
  {\bibfield  {journal} {\bibinfo  {journal} {SciPost Phys.}\ }\textbf
  {\bibinfo {volume} {2}},\ \bibinfo {pages} {016} (\bibinfo {year}
  {2017})}\BibitemShut {NoStop}%
\bibitem [{\citenamefont {Sheykhi}(2010{\natexlab{b}})}]{Sheykhi:2010wm}%
  \BibitemOpen
  \bibfield  {author} {\bibinfo {author} {\bibfnamefont {A.}~\bibnamefont
  {Sheykhi}},\ }\bibfield  {title} {\bibinfo {title} {{Entropic Corrections to
  Friedmann Equations}},\ }\href {https://doi.org/10.1103/PhysRevD.81.104011}
  {\bibfield  {journal} {\bibinfo  {journal} {Phys. Rev. D}\ }\textbf {\bibinfo
  {volume} {81}},\ \bibinfo {pages} {104011} (\bibinfo {year}
  {2010}{\natexlab{b}})}\BibitemShut {NoStop}%
\bibitem [{\citenamefont {Sheykhi}\ and\ \citenamefont
  {Hendi}(2011)}]{Sheykhi:2010yq}%
  \BibitemOpen
  \bibfield  {author} {\bibinfo {author} {\bibfnamefont {A.}~\bibnamefont
  {Sheykhi}}\ and\ \bibinfo {author} {\bibfnamefont {S.~H.}\ \bibnamefont
  {Hendi}},\ }\bibfield  {title} {\bibinfo {title} {{Power-Law Entropic
  Corrections to Newton's Law and Friedmann Equations}},\ }\href
  {https://doi.org/10.1103/PhysRevD.84.044023} {\bibfield  {journal} {\bibinfo
  {journal} {Phys. Rev. D}\ }\textbf {\bibinfo {volume} {84}},\ \bibinfo
  {pages} {044023} (\bibinfo {year} {2011})}\BibitemShut {NoStop}%
\bibitem [{\citenamefont {Cai}\ \emph {et~al.}(2010{\natexlab{a}})\citenamefont
  {Cai}, \citenamefont {Cao},\ and\ \citenamefont {Ohta}}]{Cai:2010hk}%
  \BibitemOpen
  \bibfield  {author} {\bibinfo {author} {\bibfnamefont {R.-G.}\ \bibnamefont
  {Cai}}, \bibinfo {author} {\bibfnamefont {L.-M.}\ \bibnamefont {Cao}},\ and\
  \bibinfo {author} {\bibfnamefont {N.}~\bibnamefont {Ohta}},\ }\bibfield
  {title} {\bibinfo {title} {{Friedmann Equations from Entropic Force}},\
  }\href {https://doi.org/10.1103/PhysRevD.81.061501} {\bibfield  {journal}
  {\bibinfo  {journal} {Phys. Rev. D}\ }\textbf {\bibinfo {volume} {81}},\
  \bibinfo {pages} {061501} (\bibinfo {year} {2010}{\natexlab{a}})}\BibitemShut
  {NoStop}%
\bibitem [{\citenamefont {Cai}\ and\ \citenamefont
  {Saridakis}(2011)}]{Cai:2010kp}%
  \BibitemOpen
  \bibfield  {author} {\bibinfo {author} {\bibfnamefont {Y.-F.}\ \bibnamefont
  {Cai}}\ and\ \bibinfo {author} {\bibfnamefont {E.~N.}\ \bibnamefont
  {Saridakis}},\ }\bibfield  {title} {\bibinfo {title} {{Inflation in Entropic
  Cosmology: Primordial Perturbations and non-Gaussianities}},\ }\href
  {https://doi.org/10.1016/j.physletb.2011.02.020} {\bibfield  {journal}
  {\bibinfo  {journal} {Phys. Lett. B}\ }\textbf {\bibinfo {volume} {697}},\
  \bibinfo {pages} {280} (\bibinfo {year} {2011})}\BibitemShut {NoStop}%
\bibitem [{\citenamefont {Cai}\ \emph {et~al.}(2010{\natexlab{b}})\citenamefont
  {Cai}, \citenamefont {Liu},\ and\ \citenamefont {Li}}]{Cai:2010zw}%
  \BibitemOpen
  \bibfield  {author} {\bibinfo {author} {\bibfnamefont {Y.-F.}\ \bibnamefont
  {Cai}}, \bibinfo {author} {\bibfnamefont {J.}~\bibnamefont {Liu}},\ and\
  \bibinfo {author} {\bibfnamefont {H.}~\bibnamefont {Li}},\ }\bibfield
  {title} {\bibinfo {title} {{Entropic cosmology: a unified model of inflation
  and late-time acceleration}},\ }\href
  {https://doi.org/10.1016/j.physletb.2010.05.033} {\bibfield  {journal}
  {\bibinfo  {journal} {Phys. Lett. B}\ }\textbf {\bibinfo {volume} {690}},\
  \bibinfo {pages} {213} (\bibinfo {year} {2010}{\natexlab{b}})}\BibitemShut
  {NoStop}%
\bibitem [{\citenamefont {Wei}(2010)}]{Wei:2010am}%
  \BibitemOpen
  \bibfield  {author} {\bibinfo {author} {\bibfnamefont {H.}~\bibnamefont
  {Wei}},\ }\bibfield  {title} {\bibinfo {title} {{Cosmological Constraints on
  the Modified Entropic Force Model}},\ }\href
  {https://doi.org/10.1016/j.physletb.2010.07.036} {\bibfield  {journal}
  {\bibinfo  {journal} {Phys. Lett. B}\ }\textbf {\bibinfo {volume} {692}},\
  \bibinfo {pages} {167} (\bibinfo {year} {2010})}\BibitemShut {NoStop}%
\bibitem [{\citenamefont {Sheykhi}\ and\ \citenamefont
  {Sarab}(2012)}]{Sheykhi:2012vf}%
  \BibitemOpen
  \bibfield  {author} {\bibinfo {author} {\bibfnamefont {A.}~\bibnamefont
  {Sheykhi}}\ and\ \bibinfo {author} {\bibfnamefont {K.~R.}\ \bibnamefont
  {Sarab}},\ }\bibfield  {title} {\bibinfo {title} {{Einstein Equations and
  MOND Theory from Debye Entropic Gravity}},\ }\href
  {https://doi.org/10.1088/1475-7516/2012/10/012} {\bibfield  {journal}
  {\bibinfo  {journal} {JCAP}\ }\textbf {\bibinfo {volume} {10}},\ \bibinfo
  {pages} {012}}\BibitemShut {NoStop}%
\bibitem [{\citenamefont {Komatsu}\ and\ \citenamefont
  {Kimura}(2013{\natexlab{a}})}]{Komatsu:2012zh}%
  \BibitemOpen
  \bibfield  {author} {\bibinfo {author} {\bibfnamefont {N.}~\bibnamefont
  {Komatsu}}\ and\ \bibinfo {author} {\bibfnamefont {S.}~\bibnamefont
  {Kimura}},\ }\bibfield  {title} {\bibinfo {title} {{Non-adiabatic-like
  accelerated expansion of the late universe in entropic cosmology}},\ }\href
  {https://doi.org/10.1103/PhysRevD.87.043531} {\bibfield  {journal} {\bibinfo
  {journal} {Phys. Rev. D}\ }\textbf {\bibinfo {volume} {87}},\ \bibinfo
  {pages} {043531} (\bibinfo {year} {2013}{\natexlab{a}})}\BibitemShut
  {NoStop}%
\bibitem [{\citenamefont {Komatsu}\ and\ \citenamefont
  {Kimura}(2013{\natexlab{b}})}]{Komatsu:2013qia}%
  \BibitemOpen
  \bibfield  {author} {\bibinfo {author} {\bibfnamefont {N.}~\bibnamefont
  {Komatsu}}\ and\ \bibinfo {author} {\bibfnamefont {S.}~\bibnamefont
  {Kimura}},\ }\bibfield  {title} {\bibinfo {title} {{Entropic cosmology for a
  generalized black-hole entropy}},\ }\href
  {https://doi.org/10.1103/PhysRevD.88.083534} {\bibfield  {journal} {\bibinfo
  {journal} {Phys. Rev. D}\ }\textbf {\bibinfo {volume} {88}},\ \bibinfo
  {pages} {083534} (\bibinfo {year} {2013}{\natexlab{b}})}\BibitemShut
  {NoStop}%
\bibitem [{\citenamefont {Moradpour}\ \emph {et~al.}(2018)\citenamefont
  {Moradpour}, \citenamefont {Sheykhi}, \citenamefont {Corda},\ and\
  \citenamefont {Salako}}]{Moradpour:2018Implications}%
  \BibitemOpen
  \bibfield  {author} {\bibinfo {author} {\bibfnamefont {H.}~\bibnamefont
  {Moradpour}}, \bibinfo {author} {\bibfnamefont {A.}~\bibnamefont {Sheykhi}},
  \bibinfo {author} {\bibfnamefont {C.}~\bibnamefont {Corda}},\ and\ \bibinfo
  {author} {\bibfnamefont {I.~G.}\ \bibnamefont {Salako}},\ }\bibfield  {title}
  {\bibinfo {title} {{Implications of the generalized entropy formalisms on the
  Newtonian gravity and dynamics}},\ }\href
  {https://doi.org/10.1016/j.physletb.2018.06.040} {\bibfield  {journal}
  {\bibinfo  {journal} {Phys. Lett. B}\ }\textbf {\bibinfo {volume} {783}},\
  \bibinfo {pages} {82} (\bibinfo {year} {2018})}\BibitemShut {NoStop}%
\bibitem [{\citenamefont {\c{S}enay}\ and\ \citenamefont
  {Kibaro\u{g}lu}(2019)}]{Senay:2018xaj}%
  \BibitemOpen
  \bibfield  {author} {\bibinfo {author} {\bibfnamefont {M.}~\bibnamefont
  {\c{S}enay}}\ and\ \bibinfo {author} {\bibfnamefont {S.}~\bibnamefont
  {Kibaro\u{g}lu}},\ }\bibfield  {title} {\bibinfo {title} {{$q$-Deformed
  Einstein equations from entropic force}},\ }\href
  {https://doi.org/10.1142/S0217751X18502184} {\bibfield  {journal} {\bibinfo
  {journal} {Int. J. Mod. Phys. A}\ }\textbf {\bibinfo {volume} {33}},\
  \bibinfo {pages} {1850218} (\bibinfo {year} {2019})}\BibitemShut {NoStop}%
\bibitem [{\citenamefont {Kibaro\u{g}lu}\ and\ \citenamefont
  {Senay}(2019)}]{Kibaroglu:2018mnx}%
  \BibitemOpen
  \bibfield  {author} {\bibinfo {author} {\bibfnamefont {S.}~\bibnamefont
  {Kibaro\u{g}lu}}\ and\ \bibinfo {author} {\bibfnamefont {M.}~\bibnamefont
  {Senay}},\ }\bibfield  {title} {\bibinfo {title} {{Effects of bosonic and
  fermionic $q$-deformation on the entropic gravity}},\ }\href
  {https://doi.org/10.1142/S0217732319502493} {\bibfield  {journal} {\bibinfo
  {journal} {Mod. Phys. Lett. A}\ }\textbf {\bibinfo {volume} {34}},\ \bibinfo
  {pages} {1950249} (\bibinfo {year} {2019})}\BibitemShut {NoStop}%
\bibitem [{\citenamefont {Moradpour}\ \emph {et~al.}(2019)\citenamefont
  {Moradpour}, \citenamefont {Ziaie}, \citenamefont {Ghaffari},\ and\
  \citenamefont {Feleppa}}]{Moradpour:2019GupEup}%
  \BibitemOpen
  \bibfield  {author} {\bibinfo {author} {\bibfnamefont {H.}~\bibnamefont
  {Moradpour}}, \bibinfo {author} {\bibfnamefont {A.~H.}\ \bibnamefont
  {Ziaie}}, \bibinfo {author} {\bibfnamefont {S.}~\bibnamefont {Ghaffari}},\
  and\ \bibinfo {author} {\bibfnamefont {F.}~\bibnamefont {Feleppa}},\
  }\bibfield  {title} {\bibinfo {title} {{The generalized and extended
  uncertainty principles and their implications on the Jeans mass}},\ }\href
  {https://doi.org/10.1093/mnrasl/slz098} {\bibfield  {journal} {\bibinfo
  {journal} {Mon. Not. Roy. Astron. Soc.}\ }\textbf {\bibinfo {volume} {488}},\
  \bibinfo {pages} {L69} (\bibinfo {year} {2019})}\BibitemShut {NoStop}%
\bibitem [{\citenamefont {Kibaro\u{g}lu}(2019)}]{Kibaroglu:2019dwd}%
  \BibitemOpen
  \bibfield  {author} {\bibinfo {author} {\bibfnamefont {S.}~\bibnamefont
  {Kibaro\u{g}lu}},\ }\bibfield  {title} {\bibinfo {title} {{Generalized
  entropic gravity from modified Unruh temperature}},\ }\href
  {https://doi.org/10.1142/S0217751X19501197} {\bibfield  {journal} {\bibinfo
  {journal} {Int. J. Mod. Phys. A}\ }\textbf {\bibinfo {volume} {34}},\
  \bibinfo {pages} {1950119} (\bibinfo {year} {2019})}\BibitemShut {NoStop}%
\bibitem [{\citenamefont {Kibaro\u{g}lu}\ and\ \citenamefont
  {Senay}(2020)}]{Kibaroglu:2019odt}%
  \BibitemOpen
  \bibfield  {author} {\bibinfo {author} {\bibfnamefont {S.}~\bibnamefont
  {Kibaro\u{g}lu}}\ and\ \bibinfo {author} {\bibfnamefont {M.}~\bibnamefont
  {Senay}},\ }\bibfield  {title} {\bibinfo {title} {{Friedmann equations for
  deformed entropic gravity}},\ }\href
  {https://doi.org/10.1142/S021827182050042X} {\bibfield  {journal} {\bibinfo
  {journal} {Int. J. Mod. Phys. D}\ }\textbf {\bibinfo {volume} {29}},\
  \bibinfo {pages} {2050042} (\bibinfo {year} {2020})}\BibitemShut {NoStop}%
\bibitem [{\citenamefont {Senay}(2021)}]{SENAY2021136536}%
  \BibitemOpen
  \bibfield  {author} {\bibinfo {author} {\bibfnamefont {M.}~\bibnamefont
  {Senay}},\ }\bibfield  {title} {\bibinfo {title} {Entropic gravity corrected
  by $q$-statistics, and its implications to cosmology},\ }\href
  {https://doi.org/https://doi.org/10.1016/j.physletb.2021.136536} {\bibfield
  {journal} {\bibinfo  {journal} {Physics Letters B}\ }\textbf {\bibinfo
  {volume} {820}},\ \bibinfo {pages} {136536} (\bibinfo {year}
  {2021})}\BibitemShut {NoStop}%
\bibitem [{\citenamefont {Senay}\ \emph {et~al.}(2021)\citenamefont {Senay},
  \citenamefont {Sabet},\ and\ \citenamefont {Moradpour}}]{Senay_2021}%
  \BibitemOpen
  \bibfield  {author} {\bibinfo {author} {\bibfnamefont {M.}~\bibnamefont
  {Senay}}, \bibinfo {author} {\bibfnamefont {M.~M.}\ \bibnamefont {Sabet}},\
  and\ \bibinfo {author} {\bibfnamefont {H.}~\bibnamefont {Moradpour}},\
  }\bibfield  {title} {\bibinfo {title} {Heat capacity of holographic screen
  inspires mond theory},\ }\href {https://doi.org/10.1088/1402-4896/abf618}
  {\bibfield  {journal} {\bibinfo  {journal} {Physica Scripta}\ }\textbf
  {\bibinfo {volume} {96}},\ \bibinfo {pages} {075001} (\bibinfo {year}
  {2021})}\BibitemShut {NoStop}%
\bibitem [{\citenamefont {Senay}(2024{\natexlab{a}})}]{Senay_2024}%
  \BibitemOpen
  \bibfield  {author} {\bibinfo {author} {\bibfnamefont {M.}~\bibnamefont
  {Senay}},\ }\bibfield  {title} {\bibinfo {title} {Jeans mass and gamow
  temperature: insights from $q$-deformed systems},\ }\href
  {https://doi.org/10.1088/1402-4896/ad7176} {\bibfield  {journal} {\bibinfo
  {journal} {Physica Scripta}\ }\textbf {\bibinfo {volume} {99}},\ \bibinfo
  {pages} {105001} (\bibinfo {year} {2024}{\natexlab{a}})}\BibitemShut
  {NoStop}%
\bibitem [{\citenamefont {Senay}(2024{\natexlab{b}})}]{Senay:2024Implications}%
  \BibitemOpen
  \bibfield  {author} {\bibinfo {author} {\bibfnamefont {M.}~\bibnamefont
  {Senay}},\ }\bibfield  {title} {\bibinfo {title} {Implications of q-deformed
  statistics on stellar stability},\ }\href
  {https://doi.org/https://doi.org/10.1016/j.physa.2024.130163} {\bibfield
  {journal} {\bibinfo  {journal} {Physica A}\ ,\ \bibinfo {pages} {130163}}
  (\bibinfo {year} {2024}{\natexlab{b}})}\BibitemShut {NoStop}%
\bibitem [{\citenamefont {Moradpour}\ \emph
  {et~al.}(2024{\natexlab{a}})\citenamefont {Moradpour}, \citenamefont
  {Jalalzadeh},\ and\ \citenamefont {Javaherian}}]{Moradpour:2024Fractional}%
  \BibitemOpen
  \bibfield  {author} {\bibinfo {author} {\bibfnamefont {H.}~\bibnamefont
  {Moradpour}}, \bibinfo {author} {\bibfnamefont {S.}~\bibnamefont
  {Jalalzadeh}},\ and\ \bibinfo {author} {\bibfnamefont {M.}~\bibnamefont
  {Javaherian}},\ }\bibfield  {title} {\bibinfo {title} {{Fractional stars}},\
  }\href {https://doi.org/10.1007/s10509-024-04362-y} {\bibfield  {journal}
  {\bibinfo  {journal} {Astrophys. Space Sci.}\ }\textbf {\bibinfo {volume}
  {369}},\ \bibinfo {pages} {98} (\bibinfo {year} {2024}{\natexlab{a}})},\
  \Eprint {https://arxiv.org/abs/2409.12869} {arXiv:2409.12869 [gr-qc]}
  \BibitemShut {NoStop}%
\bibitem [{\citenamefont {Moradpour}\ \emph
  {et~al.}(2024{\natexlab{b}})\citenamefont {Moradpour}, \citenamefont
  {Javaherian}, \citenamefont {Afshar},\ and\ \citenamefont
  {Jalalzadeh}}]{Moradpour:2024Tsallisian}%
  \BibitemOpen
  \bibfield  {author} {\bibinfo {author} {\bibfnamefont {H.}~\bibnamefont
  {Moradpour}}, \bibinfo {author} {\bibfnamefont {M.}~\bibnamefont
  {Javaherian}}, \bibinfo {author} {\bibfnamefont {B.}~\bibnamefont {Afshar}},\
  and\ \bibinfo {author} {\bibfnamefont {S.}~\bibnamefont {Jalalzadeh}},\
  }\bibfield  {title} {\bibinfo {title} {{Tsallisian non-extensive stars}},\
  }\href {https://doi.org/10.1016/j.physa.2024.129564} {\bibfield  {journal}
  {\bibinfo  {journal} {Physica A}\ }\textbf {\bibinfo {volume} {636}},\
  \bibinfo {pages} {129564} (\bibinfo {year} {2024}{\natexlab{b}})}\BibitemShut
  {NoStop}%
\bibitem [{\citenamefont {Kibaro\u{g}lu}\ and\ \citenamefont
  {Senay}(2025)}]{Kibaroglu:2025Anisotropic}%
  \BibitemOpen
  \bibfield  {author} {\bibinfo {author} {\bibfnamefont {S.}~\bibnamefont
  {Kibaro\u{g}lu}}\ and\ \bibinfo {author} {\bibfnamefont {M.}~\bibnamefont
  {Senay}},\ }\bibfield  {title} {\bibinfo {title} {{Anisotropic cosmology in
  q-deformed entropic gravity}},\ }\href
  {https://doi.org/10.1016/j.nuclphysb.2025.116820} {\bibfield  {journal}
  {\bibinfo  {journal} {Nucl. Phys. B}\ }\textbf {\bibinfo {volume} {1012}},\
  \bibinfo {pages} {116820} (\bibinfo {year} {2025})}\BibitemShut {NoStop}%
\bibitem [{\citenamefont {Strominger}(1993)}]{Strominger:1993si}%
  \BibitemOpen
  \bibfield  {author} {\bibinfo {author} {\bibfnamefont {A.}~\bibnamefont
  {Strominger}},\ }\bibfield  {title} {\bibinfo {title} {{Black hole
  statistics}},\ }\href {https://doi.org/10.1103/PhysRevLett.71.3397}
  {\bibfield  {journal} {\bibinfo  {journal} {Phys. Rev. Lett.}\ }\textbf
  {\bibinfo {volume} {71}},\ \bibinfo {pages} {3397} (\bibinfo {year}
  {1993})}\BibitemShut {NoStop}%
\bibitem [{\citenamefont {Arik}\ and\ \citenamefont
  {Coon}(1976)}]{Arik:1973vg}%
  \BibitemOpen
  \bibfield  {author} {\bibinfo {author} {\bibfnamefont {M.}~\bibnamefont
  {Arik}}\ and\ \bibinfo {author} {\bibfnamefont {D.~D.}\ \bibnamefont
  {Coon}},\ }\bibfield  {title} {\bibinfo {title} {{Hilbert Spaces of Analytic
  Functions and Generalized Coherent States}},\ }\href
  {https://doi.org/10.1063/1.522937} {\bibfield  {journal} {\bibinfo  {journal}
  {J. Math. Phys.}\ }\textbf {\bibinfo {volume} {17}},\ \bibinfo {pages} {524}
  (\bibinfo {year} {1976})}\BibitemShut {NoStop}%
\bibitem [{\citenamefont {Parthasarathy}\ and\ \citenamefont
  {Viswanathan}(1991)}]{Parthasarathy_1991}%
  \BibitemOpen
  \bibfield  {author} {\bibinfo {author} {\bibfnamefont {R.}~\bibnamefont
  {Parthasarathy}}\ and\ \bibinfo {author} {\bibfnamefont {K.~S.}\ \bibnamefont
  {Viswanathan}},\ }\bibfield  {title} {\bibinfo {title} {{A q-analogue of the
  supersymmetric oscillator and its q-superalgebra}},\ }\href
  {https://doi.org/10.1088/0305-4470/24/3/019} {\bibfield  {journal} {\bibinfo
  {journal} {Journal of Physics A: Mathematical and General}\ }\textbf
  {\bibinfo {volume} {24}},\ \bibinfo {pages} {613} (\bibinfo {year}
  {1991})}\BibitemShut {NoStop}%
\bibitem [{\citenamefont {Viswanathan}\ \emph {et~al.}(1992)\citenamefont
  {Viswanathan}, \citenamefont {Parthasarathy},\ and\ \citenamefont
  {Jagannathan}}]{Viswanathan_1992}%
  \BibitemOpen
  \bibfield  {author} {\bibinfo {author} {\bibfnamefont {K.~S.}\ \bibnamefont
  {Viswanathan}}, \bibinfo {author} {\bibfnamefont {R.}~\bibnamefont
  {Parthasarathy}},\ and\ \bibinfo {author} {\bibfnamefont {R.}~\bibnamefont
  {Jagannathan}},\ }\bibfield  {title} {\bibinfo {title} {{Generalized
  q-fermion oscillators and q-coherent states}},\ }\href
  {https://doi.org/10.1088/0305-4470/25/7/009} {\bibfield  {journal} {\bibinfo
  {journal} {Journal of Physics A: Mathematical and General}\ }\textbf
  {\bibinfo {volume} {25}},\ \bibinfo {pages} {L335} (\bibinfo {year}
  {1992})}\BibitemShut {NoStop}%
\bibitem [{\citenamefont {Chaichian}\ \emph {et~al.}(1993)\citenamefont
  {Chaichian}, \citenamefont {Felipe},\ and\ \citenamefont
  {Montonen}}]{Chaichian_1993}%
  \BibitemOpen
  \bibfield  {author} {\bibinfo {author} {\bibfnamefont {M.}~\bibnamefont
  {Chaichian}}, \bibinfo {author} {\bibfnamefont {R.~G.}\ \bibnamefont
  {Felipe}},\ and\ \bibinfo {author} {\bibfnamefont {C.}~\bibnamefont
  {Montonen}},\ }\bibfield  {title} {\bibinfo {title} {{Statistics of
  q-oscillators, quons and relations to fractional statistics}},\ }\href
  {https://doi.org/10.1088/0305-4470/26/16/018} {\bibfield  {journal} {\bibinfo
   {journal} {Journal of Physics A: Mathematical and General}\ }\textbf
  {\bibinfo {volume} {26}},\ \bibinfo {pages} {4017} (\bibinfo {year}
  {1993})}\BibitemShut {NoStop}%
\bibitem [{\citenamefont {Ubriaco}(1997)}]{Ubriaco1997}%
  \BibitemOpen
  \bibfield  {author} {\bibinfo {author} {\bibfnamefont {M.~R.}\ \bibnamefont
  {Ubriaco}},\ }\bibfield  {title} {\bibinfo {title} {Anyonic behavior of
  quantum group gases},\ }\href {https://doi.org/10.1103/PhysRevE.55.291}
  {\bibfield  {journal} {\bibinfo  {journal} {Phys. Rev. E}\ }\textbf {\bibinfo
  {volume} {55}},\ \bibinfo {pages} {291} (\bibinfo {year} {1997})}\BibitemShut
  {NoStop}%
\bibitem [{\citenamefont {Lavagno}\ and\ \citenamefont
  {Swamy}(2002)}]{LAVAGNO2002310}%
  \BibitemOpen
  \bibfield  {author} {\bibinfo {author} {\bibfnamefont {A.}~\bibnamefont
  {Lavagno}}\ and\ \bibinfo {author} {\bibfnamefont {P.}~\bibnamefont
  {Swamy}},\ }\bibfield  {title} {\bibinfo {title} {{$q$-Deformed structures
  and nonextensive statistics: a comparative study}},\ }\href
  {https://doi.org/https://doi.org/10.1016/S0378-4371(01)00680-X} {\bibfield
  {journal} {\bibinfo  {journal} {Physica A: Statistical Mechanics and its
  Applications}\ }\textbf {\bibinfo {volume} {305}},\ \bibinfo {pages} {310}
  (\bibinfo {year} {2002})}\BibitemShut {NoStop}%
\bibitem [{\citenamefont {Algin}\ and\ \citenamefont
  {Senay}(2012)}]{Algin:2012df}%
  \BibitemOpen
  \bibfield  {author} {\bibinfo {author} {\bibfnamefont {A.}~\bibnamefont
  {Algin}}\ and\ \bibinfo {author} {\bibfnamefont {M.}~\bibnamefont {Senay}},\
  }\bibfield  {title} {\bibinfo {title} {{High temperature behavior of a
  deformed Fermi gas obeying interpolating statistics}},\ }\href
  {https://doi.org/10.1103/PhysRevE.85.041123} {\bibfield  {journal} {\bibinfo
  {journal} {Phys. Rev. E}\ }\textbf {\bibinfo {volume} {85}},\ \bibinfo
  {pages} {041123} (\bibinfo {year} {2012})}\BibitemShut {NoStop}%
\bibitem [{\citenamefont {Algin}\ and\ \citenamefont
  {Senay}(2016{\natexlab{a}})}]{Algin:2016cuo}%
  \BibitemOpen
  \bibfield  {author} {\bibinfo {author} {\bibfnamefont {A.}~\bibnamefont
  {Algin}}\ and\ \bibinfo {author} {\bibfnamefont {M.}~\bibnamefont {Senay}},\
  }\bibfield  {title} {\bibinfo {title} {{Fermionic q -deformation and its
  connection to thermal effective mass of a quasiparticle}},\ }\href
  {https://doi.org/10.1016/j.physa.2015.12.014} {\bibfield  {journal} {\bibinfo
   {journal} {Physica A}\ }\textbf {\bibinfo {volume} {447}},\ \bibinfo {pages}
  {232} (\bibinfo {year} {2016}{\natexlab{a}})}\BibitemShut {NoStop}%
\bibitem [{\citenamefont {Algin}\ and\ \citenamefont
  {Senay}(2016{\natexlab{b}})}]{Algin:2016df}%
  \BibitemOpen
  \bibfield  {author} {\bibinfo {author} {\bibfnamefont {A.}~\bibnamefont
  {Algin}}\ and\ \bibinfo {author} {\bibfnamefont {M.}~\bibnamefont {Senay}},\
  }\bibfield  {title} {\bibinfo {title} {{General thermostatistical properties
  of a $q$-deformed fermion gas in two dimensions}},\ }\href
  {https://doi.org/10.1088/1742-6596/766/1/012008} {\bibfield  {journal}
  {\bibinfo  {journal} {Journal of Physics: Conference Series}\ }\textbf
  {\bibinfo {volume} {766}},\ \bibinfo {pages} {012008} (\bibinfo {year}
  {2016}{\natexlab{b}})}\BibitemShut {NoStop}%
\bibitem [{\citenamefont {Algin}\ \emph {et~al.}(2015)\citenamefont {Algin},
  \citenamefont {Irk},\ and\ \citenamefont {Topcu}}]{Algin2015}%
  \BibitemOpen
  \bibfield  {author} {\bibinfo {author} {\bibfnamefont {A.}~\bibnamefont
  {Algin}}, \bibinfo {author} {\bibfnamefont {D.}~\bibnamefont {Irk}},\ and\
  \bibinfo {author} {\bibfnamefont {G.}~\bibnamefont {Topcu}},\ }\bibfield
  {title} {\bibinfo {title} {Anyonic behavior of an intermediate-statistics
  fermion gas model},\ }\href {https://doi.org/10.1103/PhysRevE.91.062131}
  {\bibfield  {journal} {\bibinfo  {journal} {Phys. Rev. E}\ }\textbf {\bibinfo
  {volume} {91}},\ \bibinfo {pages} {062131} (\bibinfo {year}
  {2015})}\BibitemShut {NoStop}%
\bibitem [{\citenamefont {Algin}\ \emph {et~al.}(2017)\citenamefont {Algin},
  \citenamefont {Arik}, \citenamefont {Senay},\ and\ \citenamefont
  {Topcu}}]{Algin2017}%
  \BibitemOpen
  \bibfield  {author} {\bibinfo {author} {\bibfnamefont {A.}~\bibnamefont
  {Algin}}, \bibinfo {author} {\bibfnamefont {M.}~\bibnamefont {Arik}},
  \bibinfo {author} {\bibfnamefont {M.}~\bibnamefont {Senay}},\ and\ \bibinfo
  {author} {\bibfnamefont {G.}~\bibnamefont {Topcu}},\ }\bibfield  {title}
  {\bibinfo {title} {Thermostatistics of bosonic and fermionic fibonacci
  oscillators},\ }\href {https://doi.org/10.1142/S0217979216502477} {\bibfield
  {journal} {\bibinfo  {journal} {International Journal of Modern Physics B}\
  }\textbf {\bibinfo {volume} {31}},\ \bibinfo {pages} {1650247} (\bibinfo
  {year} {2017})}\BibitemShut {NoStop}%
\bibitem [{\citenamefont {Mohammadzadeh}\ \emph {et~al.}(2017)\citenamefont
  {Mohammadzadeh}, \citenamefont {Azizian-Kalandaragh}, \citenamefont
  {Cheraghpour},\ and\ \citenamefont {Adli}}]{Mohammadzadeh_2017}%
  \BibitemOpen
  \bibfield  {author} {\bibinfo {author} {\bibfnamefont {H.}~\bibnamefont
  {Mohammadzadeh}}, \bibinfo {author} {\bibfnamefont {Y.}~\bibnamefont
  {Azizian-Kalandaragh}}, \bibinfo {author} {\bibfnamefont {N.}~\bibnamefont
  {Cheraghpour}},\ and\ \bibinfo {author} {\bibfnamefont {F.}~\bibnamefont
  {Adli}},\ }\bibfield  {title} {\bibinfo {title} {Thermodynamic geometry,
  condensation and debye model of two-parameter deformed statistics},\ }\href
  {https://doi.org/10.1088/1742-5468/aa7ee0} {\bibfield  {journal} {\bibinfo
  {journal} {Journal of Statistical Mechanics: Theory and Experiment}\ }\textbf
  {\bibinfo {volume} {2017}},\ \bibinfo {pages} {083104} (\bibinfo {year}
  {2017})}\BibitemShut {NoStop}%
\bibitem [{\citenamefont {Nutku}\ \emph {et~al.}(2019)\citenamefont {Nutku},
  \citenamefont {Sen},\ and\ \citenamefont {Aydiner}}]{NUTKU2019122041}%
  \BibitemOpen
  \bibfield  {author} {\bibinfo {author} {\bibfnamefont {F.}~\bibnamefont
  {Nutku}}, \bibinfo {author} {\bibfnamefont {K.}~\bibnamefont {Sen}},\ and\
  \bibinfo {author} {\bibfnamefont {E.}~\bibnamefont {Aydiner}},\ }\bibfield
  {title} {\bibinfo {title} {Complexity study of q-deformed quantum harmonic
  oscillator},\ }\href
  {https://doi.org/https://doi.org/10.1016/j.physa.2019.122041} {\bibfield
  {journal} {\bibinfo  {journal} {Physica A: Statistical Mechanics and its
  Applications}\ }\textbf {\bibinfo {volume} {533}},\ \bibinfo {pages} {122041}
  (\bibinfo {year} {2019})}\BibitemShut {NoStop}%
\bibitem [{\citenamefont {Altintas}\ \emph {et~al.}(2020)\citenamefont
  {Altintas}, \citenamefont {Ozaydin},\ and\ \citenamefont
  {Bay\i{}nd\i{}r}}]{Altintas:2020eqx}%
  \BibitemOpen
  \bibfield  {author} {\bibinfo {author} {\bibfnamefont {A.~A.}\ \bibnamefont
  {Altintas}}, \bibinfo {author} {\bibfnamefont {F.}~\bibnamefont {Ozaydin}},\
  and\ \bibinfo {author} {\bibfnamefont {C.}~\bibnamefont {Bay\i{}nd\i{}r}},\
  }\bibfield  {title} {\bibinfo {title} {{$q$-Deformed three-level quantum
  logic}},\ }\href {https://doi.org/10.1007/s11128-020-02755-w} {\bibfield
  {journal} {\bibinfo  {journal} {Quant. Inf. Proc.}\ }\textbf {\bibinfo
  {volume} {19}},\ \bibinfo {pages} {247} (\bibinfo {year} {2020})}\BibitemShut
  {NoStop}%
\bibitem [{\citenamefont {Ozaydin}\ \emph {et~al.}(2023)\citenamefont
  {Ozaydin}, \citenamefont {M\"ustecapl\i{}o\u{g}lu},\ and\ \citenamefont
  {Hakio\u{g}lu}}]{Ozaydin:2023ege}%
  \BibitemOpen
  \bibfield  {author} {\bibinfo {author} {\bibfnamefont {F.}~\bibnamefont
  {Ozaydin}}, \bibinfo {author} {\bibfnamefont {O.~E.}\ \bibnamefont
  {M\"ustecapl\i{}o\u{g}lu}},\ and\ \bibinfo {author} {\bibfnamefont
  {T.}~\bibnamefont {Hakio\u{g}lu}},\ }\bibfield  {title} {\bibinfo {title}
  {{Powering quantum Otto engines only with $q$-deformation of the working
  substance}},\ }\href {https://doi.org/10.1103/PhysRevE.108.054103} {\bibfield
   {journal} {\bibinfo  {journal} {Phys. Rev. E}\ }\textbf {\bibinfo {volume}
  {108}},\ \bibinfo {pages} {054103} (\bibinfo {year} {2023})}\BibitemShut
  {NoStop}%
\bibitem [{\citenamefont {Marinho}\ \emph {et~al.}(2024)\citenamefont
  {Marinho}, \citenamefont {Brito}, \citenamefont {Viswanathan},\ and\
  \citenamefont {Bezerra}}]{Marinho2024}%
  \BibitemOpen
  \bibfield  {author} {\bibinfo {author} {\bibfnamefont {A.~A.}\ \bibnamefont
  {Marinho}}, \bibinfo {author} {\bibfnamefont {F.~A.}\ \bibnamefont {Brito}},
  \bibinfo {author} {\bibfnamefont {G.~M.}\ \bibnamefont {Viswanathan}},\ and\
  \bibinfo {author} {\bibfnamefont {C.~G.}\ \bibnamefont {Bezerra}},\
  }\bibfield  {title} {\bibinfo {title} {{Intermediate statistics: Addressing
  the thermoelectric properties of solids}},\ }\href
  {https://doi.org/10.1103/PhysRevE.110.034136} {\bibfield  {journal} {\bibinfo
   {journal} {Phys. Rev. E}\ }\textbf {\bibinfo {volume} {110}},\ \bibinfo
  {pages} {034136} (\bibinfo {year} {2024})}\BibitemShut {NoStop}%
\bibitem [{\citenamefont {Boumali}\ \emph {et~al.}(2023)\citenamefont
  {Boumali}, \citenamefont {Bouzenada}, \citenamefont {Zare},\ and\
  \citenamefont {Hassanabadi}}]{BOUMALI2023129134}%
  \BibitemOpen
  \bibfield  {author} {\bibinfo {author} {\bibfnamefont {A.}~\bibnamefont
  {Boumali}}, \bibinfo {author} {\bibfnamefont {A.}~\bibnamefont {Bouzenada}},
  \bibinfo {author} {\bibfnamefont {S.}~\bibnamefont {Zare}},\ and\ \bibinfo
  {author} {\bibfnamefont {H.}~\bibnamefont {Hassanabadi}},\ }\bibfield
  {title} {\bibinfo {title} {Thermal properties of the $q$-deformed spin-one
  dkp oscillator},\ }\href
  {https://doi.org/https://doi.org/10.1016/j.physa.2023.129134} {\bibfield
  {journal} {\bibinfo  {journal} {Physica A: Statistical Mechanics and its
  Applications}\ }\textbf {\bibinfo {volume} {628}},\ \bibinfo {pages} {129134}
  (\bibinfo {year} {2023})}\BibitemShut {NoStop}%
\bibitem [{\citenamefont {Boutabba}\ \emph {et~al.}(2022)\citenamefont
  {Boutabba}, \citenamefont {Grira},\ and\ \citenamefont
  {Eleuch}}]{Boutabba2022}%
  \BibitemOpen
  \bibfield  {author} {\bibinfo {author} {\bibfnamefont {N.}~\bibnamefont
  {Boutabba}}, \bibinfo {author} {\bibfnamefont {S.}~\bibnamefont {Grira}},\
  and\ \bibinfo {author} {\bibfnamefont {H.}~\bibnamefont {Eleuch}},\
  }\bibfield  {title} {\bibinfo {title} {Analysis of a $q$-deformed hyperbolic
  short laser pulse in a multi-level atomic system},\ }\href
  {https://doi.org/10.1038/s41598-022-13407-7} {\bibfield  {journal} {\bibinfo
  {journal} {Scientific Reports}\ }\textbf {\bibinfo {volume} {12}},\ \bibinfo
  {pages} {9308} (\bibinfo {year} {2022})}\BibitemShut {NoStop}%
\bibitem [{\citenamefont {Nutku}\ and\ \citenamefont
  {Aydiner}(2018)}]{Nutku_2018}%
  \BibitemOpen
  \bibfield  {author} {\bibinfo {author} {\bibfnamefont {F.}~\bibnamefont
  {Nutku}}\ and\ \bibinfo {author} {\bibfnamefont {E.}~\bibnamefont
  {Aydiner}},\ }\bibfield  {title} {\bibinfo {title} {Investigation of
  bose-einstein condensates in $q$-deformed potentials with first order
  perturbation theory},\ }\href {https://doi.org/10.1088/0253-6102/69/2/154}
  {\bibfield  {journal} {\bibinfo  {journal} {Communications in Theoretical
  Physics}\ }\textbf {\bibinfo {volume} {69}},\ \bibinfo {pages} {154}
  (\bibinfo {year} {2018})}\BibitemShut {NoStop}%
\bibitem [{\citenamefont {Marinho}\ \emph {et~al.}(2014)\citenamefont
  {Marinho}, \citenamefont {Brito},\ and\ \citenamefont
  {Chesman}}]{MARINHO201474}%
  \BibitemOpen
  \bibfield  {author} {\bibinfo {author} {\bibfnamefont {A.~A.}\ \bibnamefont
  {Marinho}}, \bibinfo {author} {\bibfnamefont {F.~A.}\ \bibnamefont {Brito}},\
  and\ \bibinfo {author} {\bibfnamefont {C.}~\bibnamefont {Chesman}},\
  }\bibfield  {title} {\bibinfo {title} {Fibonacci oscillators in the landau
  diamagnetism problem},\ }\href
  {https://doi.org/https://doi.org/10.1016/j.physa.2014.06.008} {\bibfield
  {journal} {\bibinfo  {journal} {Physica A: Statistical Mechanics and its
  Applications}\ }\textbf {\bibinfo {volume} {411}},\ \bibinfo {pages} {74}
  (\bibinfo {year} {2014})}\BibitemShut {NoStop}%
\bibitem [{\citenamefont {Gavrilik}\ and\ \citenamefont
  {Mishchenko}(2014)}]{Gavrilik2014}%
  \BibitemOpen
  \bibfield  {author} {\bibinfo {author} {\bibfnamefont {A.~M.}\ \bibnamefont
  {Gavrilik}}\ and\ \bibinfo {author} {\bibfnamefont {Y.~A.}\ \bibnamefont
  {Mishchenko}},\ }\bibfield  {title} {\bibinfo {title} {Virial coefficients in
  the $(\stackrel{\ifmmode \tilde{}\else
  \~{}\fi{}}{\ensuremath{\mu}},q)$-deformed bose gas model related to
  compositeness of particles and their interaction: Temperature-dependence
  problem},\ }\href {https://doi.org/10.1103/PhysRevE.90.052147} {\bibfield
  {journal} {\bibinfo  {journal} {Phys. Rev. E}\ }\textbf {\bibinfo {volume}
  {90}},\ \bibinfo {pages} {052147} (\bibinfo {year} {2014})}\BibitemShut
  {NoStop}%
\bibitem [{\citenamefont {Ismail}\ and\ \citenamefont
  {Stanton}(2003)}]{ISMAIL2003259}%
  \BibitemOpen
  \bibfield  {author} {\bibinfo {author} {\bibfnamefont {M.~E.}\ \bibnamefont
  {Ismail}}\ and\ \bibinfo {author} {\bibfnamefont {D.}~\bibnamefont
  {Stanton}},\ }\bibfield  {title} {\bibinfo {title} {Applications of
  $q$-taylor theorems},\ }\href
  {https://doi.org/https://doi.org/10.1016/S0377-0427(02)00644-1} {\bibfield
  {journal} {\bibinfo  {journal} {Journal of Computational and Applied
  Mathematics}\ }\textbf {\bibinfo {volume} {153}},\ \bibinfo {pages} {259}
  (\bibinfo {year} {2003})}\BibitemShut {NoStop}%
\bibitem [{\citenamefont {Jalalzadeh}\ \emph {et~al.}(2023)\citenamefont
  {Jalalzadeh}, \citenamefont {Moradpour},\ and\ \citenamefont
  {Moniz}}]{JALALZADEH2023101320}%
  \BibitemOpen
  \bibfield  {author} {\bibinfo {author} {\bibfnamefont {S.}~\bibnamefont
  {Jalalzadeh}}, \bibinfo {author} {\bibfnamefont {H.}~\bibnamefont
  {Moradpour}},\ and\ \bibinfo {author} {\bibfnamefont {P.}~\bibnamefont
  {Moniz}},\ }\bibfield  {title} {\bibinfo {title} {Modified cosmology from
  quantum deformed entropy},\ }\href
  {https://doi.org/https://doi.org/10.1016/j.dark.2023.101320} {\bibfield
  {journal} {\bibinfo  {journal} {Physics of the Dark Universe}\ }\textbf
  {\bibinfo {volume} {42}},\ \bibinfo {pages} {101320} (\bibinfo {year}
  {2023})}\BibitemShut {NoStop}%
\bibitem [{\citenamefont {Wald}(1984)}]{Wald:1984GeneralRelativity}%
  \BibitemOpen
  \bibfield  {author} {\bibinfo {author} {\bibfnamefont {R.~M.}\ \bibnamefont
  {Wald}},\ }\href {https://doi.org/10.7208/chicago/9780226870373.001.0001}
  {\emph {\bibinfo {title} {{General Relativity}}}}\ (\bibinfo  {publisher}
  {Chicago Univ. Pr.},\ \bibinfo {address} {Chicago, USA},\ \bibinfo {year}
  {1984})\BibitemShut {NoStop}%
\bibitem [{\citenamefont {Unruh}(1976)}]{Unruh:1976Notes}%
  \BibitemOpen
  \bibfield  {author} {\bibinfo {author} {\bibfnamefont {W.~G.}\ \bibnamefont
  {Unruh}},\ }\bibfield  {title} {\bibinfo {title} {{Notes on black hole
  evaporation}},\ }\href {https://doi.org/10.1103/PhysRevD.14.870} {\bibfield
  {journal} {\bibinfo  {journal} {Phys. Rev. D}\ }\textbf {\bibinfo {volume}
  {14}},\ \bibinfo {pages} {870} (\bibinfo {year} {1976})}\BibitemShut
  {NoStop}%
\bibitem [{\citenamefont {Brans}\ and\ \citenamefont
  {Dicke}(1961)}]{Brans:1961sx}%
  \BibitemOpen
  \bibfield  {author} {\bibinfo {author} {\bibfnamefont {C.}~\bibnamefont
  {Brans}}\ and\ \bibinfo {author} {\bibfnamefont {R.~H.}\ \bibnamefont
  {Dicke}},\ }\bibfield  {title} {\bibinfo {title} {{Mach's principle and a
  relativistic theory of gravitation}},\ }\href
  {https://doi.org/10.1103/PhysRev.124.925} {\bibfield  {journal} {\bibinfo
  {journal} {Phys. Rev.}\ }\textbf {\bibinfo {volume} {124}},\ \bibinfo {pages}
  {925} (\bibinfo {year} {1961})}\BibitemShut {NoStop}%
\end{thebibliography}%

\end{document}